\documentclass[11pt,a4paper]{article}
\usepackage[english]{babel}
\usepackage{a4wide}	    % Layout - smaller margins
\usepackage{pdflscape}  % Rotate pdf landscape pages
\usepackage[scaled=1, helvratio=0.95]{newtxtext}       % Better text fonts
\usepackage{newtxmath}       % Better math fonts
\usepackage[utf8]{inputenc}
\usepackage{fix-cm}          % Overrides restrictions on CM and EC fonts
\usepackage[T1]{fontenc}    % The font encoding used. Always load after the last font package.

\usepackage[ left=2.00cm, right=2.00cm, top=3.00cm, bottom=3.00cm]{geometry}
\usepackage{appendix}
\usepackage{amsmath}

\usepackage{amsthm}
\usepackage{mathtools}

\newtheorem{definition}{Definition}
\newtheorem{property}{Property}

\mathtoolsset{showonlyrefs}	% Show only labels if make a reference to it
\usepackage{amsmath}
\usepackage{xspace}  % add space when needed using \xspace
\usepackage[hidelinks]{hyperref}

\makeatletter
\newcommand{\ALC@uniqueautorefname}{line}
\makeatother
\addto\extrasenglish{  % \autoref don't capalize in this class so don't use or change defs

}

\usepackage{enumitem}	% Better lists
\usepackage{pdflscape}  % Rotate pdf landscape pages

\usepackage{tabularx}
\usepackage{booktabs}	% Pretty tables
\usepackage{graphicx}  % loaded by the class
\usepackage{subcaption}

\usepackage{tikz}
\usepackage{pgfplots}
\usepgfplotslibrary{fillbetween}
\usetikzlibrary{patterns}
\usetikzlibrary{arrows.meta}
\pgfplotsset{compat=1.15}

\usepackage{algorithm}
\usepackage{algorithmic}
\usepackage{braket}
\newcolumntype{Z}{>{\raggedleft\arraybackslash}X}
\newcolumntype{Y}{>{\centering\arraybackslash}X}
\newcolumntype{C}{>{\footnotesize}c}
\newcolumntype{R}{>{\footnotesize}r}
\newcolumntype{L}{>{\footnotesize}l}
\newcolumntype{P}[1]{>{\footnotesize\raggedright}b{#1}}

\usepackage[round, sort&compress, longnamesfirst]{natbib}   % Natbib for better citations use \citet and \citep
\usepackage{enumitem}	% Customizing lists 

\newcommand{\R}{\mathbb{R}}
\newcommand{\Rp}{\mathbb{R}^p}

\newcommand{\Rpgeq}{\mathbb{R}^p_{\geqq}}

\newcommand{\YN}{\mathcal{Y}_N}
\newcommand{\Y}{\mathcal{Y}}
\newcommand{\X}{\mathcal{X}}
\newcommand{\UB}{\mathcal{U}}
\newcommand{\LB}{\mathcal{L}}
\newcommand{\NU}{\mathcal{N}(\UB)}

\newcommand{\nodeList}{\mathcal{T}}
\usepackage{bbm}

\makeatletter
\def\@maketitle{%
  \newpage
  \null
  \vskip 2em%
  \let \footnote \thanks
	{\Large\bfseries\noindent\@title\par}%
	\vskip 1.5em%
	{\large
	  \lineskip .5em%
		\raggedright\@author}
	\vskip 1em%
	{\noindent\small\@date}%
  \par
  \vskip 1.5em}
\makeatother

\renewenvironment{abstract}{\parindent0pt\textbf{Abstract:}}{}
\usepackage[
authormarkup=none, 
todonotes={textsize=scriptsize}
]{changes} % Use \usepackage[final]{changes} for keeping the last changes. 
\definechangesauthor[name={Sophie}, color=green]{SP}
\definechangesauthor[name={Nicolas}, color=orange]{NF}

\usepackage{tabu}
\usepackage{multirow}
\begin{document}
	
%% ---------------------------------------------------------------------
%% TexStudio options
% !TeX spellcheck = en_GB	
% !TeX encoding = utf8
% !TeX root = main.tex
%% ---------------------------------------------------------------------

%% ---------------------------------------------------------------------
%% Title, authors and abstract
%% ---------------------------------------------------------------------
\title{Enhancing Branch-and-Bound for Multi-Objective 0-1 Programming}
% Variable fixing for multi-objective branch-and-bound
% and node selection 

\author{
Nicolas Forget\thanks{Corresponding author (\url{nforget@econ.au.dk}).}\\
%,\; Sune Lauth Gadegaard, Lars Relund Nielsen\\
{\small Department of Economics and Business Economics, School of Business and Social Sciences, Aarhus University, Fuglesangs All\'{e} 4, DK-8210 Aarhus V, Denmark}\\[0.3em]
Sophie N. Parragh\\
{\small Institute of Production and Logistics Management, Johannes Kepler University Linz, Altenberger Stra{\ss}e 69, 4040 Linz, Austria}\\[0.3em]
}

\date{\today}

\maketitle

\begin{abstract}

In the bi-objective branch-and-bound literature, a key ingredient is objective branching, i.e. to create smaller and disjoint sub-problems in the objective space, obtained from the partial dominance of the lower bound set by the upper bound set. When considering three or more objective functions, however, applying objective branching becomes more complex, and its benefit has so far been unclear. In this paper, we investigate several ingredients which allow to better exploit objective branching in a multi-objective setting. We extend the idea of probing to multiple objectives, enhance it in several ways, and show that when coupled with objective branching, it results in significant speed-ups in terms of CPU times. We also investigate cut generation based on the objective branching constraints. Besides, we generalize the best-bound idea for node selection to multiple objectives and we show that the proposed rules outperform the, in the multi-objective literature, commonly employed depth-first and breadth-first strategies. We also analyze problem specific branching rules. We test the proposed ideas on available benchmark instances for three problem classes with three and four objectives, namely the capacitated facility location problem, the uncapacitated facility location problem, and the knapsack problem. Our enhanced multi-objective branch-and-bound algorithm outperforms the best existing branch-and-bound based approach and is the first to obtain competitive and even slightly better results than a state-of-the-art objective space search method on a subset of the problem classes.

\vskip 1 em

{\em Keywords:}

\end{abstract}
%% ---------------------------------------------------------------------
\section{Introduction}

In many real-world problem situations, decision makers have to consider %may be 
%interested in optimizing 
several different objectives simultaneously, such as, e.g., %minimizing 
travel times, %minimizing travelling 
costs, 
and %minimizing 
CO2 emissions. %, ect. 
These objectives are often conflicting, which means  %meaning 
that the optimal solution for one of the objectives is often not optimal for the others. Instead, one may search for all the 
optimal trade-off solutions. 
%interesting trade-offs between the objectives. 
For this purpose, a multi-objective optimization problem is solved. We focus here on solving \emph{Multi-Objective Integer Linear Problems} (MOILP). %\emph{Multi-Objective Combinatorial Problems} (MOCO).

A MOILP can be solved using an \emph{Objective Space Search} (OSS) algorithm, which consists of solving a series of single-objective %combinatorial 
problems obtained by \emph{scalarizing} the objective functions so that all optimal  %interesting 
trade-offs are enumerated \citep{Ehrgott05}. The main advantage of this methodology is that the power of single-objective solvers can be used. Consequently, objective space search algorithms have received much attention over the past decades (see e.g., \cite{UlunguTeghem1995}; \cite{Visee95}; \cite{SylvaCremaDisjunction}; \cite{OzlenRecursion}; \cite{KirlikSayinTwoStageScalarazation}; \cite{BolandQuadrantShrinking}; \cite{BolandLShaped}; \cite{Tamby20}). %\comment[id=NF]{Introduction or related works?}

Alternatively, a MOILP can be solved using a \emph{Decision Space Search} (DSS) algorithm, typically a \emph{Multi-Objective Branch \& Bound} (MOBB) algorithm. 
Almost all recent contributions in this area address the bi-objective case \citep{StidsenAndersenDammann2014,Gadegaard2019,parragh2019branch,adelgren2021branch}, and they all %All of them 
rely on an efficient branching scheme that creates sub-problems in the objective space. \citet{Forget20a} have recently generalized this scheme to problems with more than two objectives.
However, in its straightforward form, the speed-ups observed for the two-objective case did not translate to the three or more objective case.
In this paper, we take \citet{Forget20a}'s work as the starting point and of we propose an enhanced MOBB framework designed to solve MOILP with three or more objective functions. %First, the framework relies on an efficient branching scheme that creates sub-problems in the objective space %\citep{StidsenAndersenDammann2014,Gadegaard2019,parragh2019branch,Forget20a}. 
In order to improve the performance of the MOBB, 
we generalize the idea of probing to the multi-objective case, which allows us to better exploit the constraints generated by the branching scheme. Probing is a technique 
successfully employed in single objective branch-and-bound to locally reduce the domains of the decision variables. In a 0-1 integer context, this results in fixing variables to either $0$ or $1$ \citep{savelsbergh1994preprocessing}. To the best of our knowledge, probing has not been generalized to more than two objectives.
%we exploit the constraints generated by the branching scheme to efficiently perform probing, a technique used in the single-objective case to fix variables to certain values.
Moreover, we investigate whether a decrease in CPU times can be achieved by deriving stronger %generating 
cuts from %based on 
the constraints generated by  %of 
the objective branching scheme. Then, we investigate new node selection rules based on the best-bound principle, and compare them to the traditional breadth and depth-first strategies typically used in the MOBB literature. Finally, we show through a computational study that the suggested improvements lead to a significant speed-up for the framework in terms of CPU time, and that our algorithm is competitive with recent OSS algorithms from the literature on some of the problem classes considered in this paper. %\comment[id=NF]{Let's see when the results come out}

The paper is organized as follows. In Section~\ref{sec:def}, we present the notation and definitions used throughout the paper. In Section~\ref{sec:relatedWorks}, we discuss %provide a review of 
related work, and in Section~\ref{sec:BB}, we describe the basic MOBB framework used here. Sections~\ref{sec:OB} and \ref{sec:nodeSel} are dedicated to the main novelties of our framework, namely variable fixing, cut generation, and node selection rules. Finally, in Section~\ref{sec:expe}, we present the computational study, and %highlight 
our conclusions in Section~\ref{sec:conclu}.

%Hence, in this paper, we investigate how objective branching constraints can be further utilized. In particular, we:

%\begin{itemize}
%    \item propose a way to efficiently fix some variables to one of their bounds during the resolution
%    \item discuss cuts derived from objective branching constraints
%    \item show on a set of three problem classes that the approach performs very well when coupled with objective branching
%\end{itemize}

%Moreover, we explore simple node selection rules that are based on the best-bound principle. In particular, we:

%\begin{itemize}
%    \item discuss a node selection rule based on weighted-sum scalarizations
%    \item present an alternative rule based on gap measures
%    \item report results regarding the efficiency of these rules compared to the classical depth-first and breadth-first strategies on a set of three problem classes
%\end{itemize}

%lead to an efficient framework in the bi-objective case, but it seems not to be sufficient by itself when three or more objective functions are considered. Hence, in this paper, we aim at ...

%  Indeed, if it proved to be beneficial for some problem classes, it led to larger CPU times for others.

\section{Definitions and notation}
\label{sec:def}

A MOILP with $n$ variables, $p$ objectives, and $m$ constraints is written as follows:
\begin{equation}
    P: \quad \min\{z(x):x\in \mathcal{X}\}
\end{equation}
where the $p$ objective functions $z(x) = Cx$ are defined by a $p \times n$ matrix of objective coefficients $C$. The \emph{feasible set} $\X = \{x \in \{0,1\}: Ax \geqq b\}$ is given by a $m \times n$ matrix of constraint coefficients $A$, and a right-hand-side vector $b$ of size $m$. The image of the feasible set in the objective space is $\Y := C\X = \{Cx : x\in \X\}$.

Since the objective function is vector-valued, the following %new 
operators are %need to be 
introduced to compare solutions. Let $y^1, y^2 \in \Y$, $y^1$ \emph{weakly dominates} $y^2$ ($y^1 \leqq y^2$) if $y^1_k \leq y^2_k$ for all $k = 1,...,p$. Besides, $y^1$ \emph{dominates} $y^2$ ($y^1 \leqslant y^2$) if $y^1 \leqq y^2$ and $y^1 \ne y^2$. These relations also extend to sets of points. Let $\mathcal{S}^1, \mathcal{S}^2 \subset \R^p$, we say that $\mathcal{S}^1$ \emph{dominates} $\mathcal{S}^2$ if for all $s^2 \in \mathcal{S}^2$, there exists $s^1 \in \mathcal{S}^1$ such that $s^1 \leqslant s^2$. The set $\mathcal{S}^1$ \emph{partially dominates} $\mathcal{S}^2$ if $\mathcal{S}^1$ does not dominate $\mathcal{S}^2$, but there is at least one $s^2 \in \mathcal{S}^2$ such that there exists $s^1 \in \mathcal{S}^1$ such that $s^1 \leqslant s^2$.

Derived from the dominance relations, the \emph{set of non-dominated points} is defined as $\YN = \{y \in \Y : \nexists y' \in \Y, y'\leqslant y\}$. This notation can be extended to any set $\mathcal{S} \subset \R^p$, i.e. $\mathcal{S}_N = \{y \in \mathcal{S} : \nexists y' \in \mathcal{S}, y'\leqslant y\}$. Moreover, we define the set $\Rpgeq = \{y \in \Rp : y \geqq 0\}$.

\cite{Ehrgott07} introduced the notions of lower and upper bound sets for $\mathcal{S}_N$, $\mathcal{S} \subset \R^p$, and extended the concept of lower and upper bound to the multi-objective case. Let $\mathcal{S}^1, \mathcal{S}^2 \subset \Rpgeq$, we define the operation $\mathcal{S}^1 + \mathcal{S}^2$ as the Minkowski sum, i.e., $\mathcal{S}^1 + \mathcal{S}^2 = \{s^1 + s^2 : s^1\in \mathcal{S}^1, s^2 \in \mathcal{S}^2\}$. Moreover, $\mathcal{S}^1$ is $\Rp$-closed if $\mathcal{S}^1 + \Rp$ is closed, and $\Rp$-bounded if there exists $s \in \Rp$ such that $\mathcal{S}^1 \subset \{s\} + \Rpgeq$. The definition of \cite{Ehrgott07} is recalled in Definition~\ref{def:boundSets}.

\begin{definition}
    \citep{Ehrgott07} Let $\mathcal{S} \subset \R^p$.
    \begin{itemize}
        \item A lower bound set $\LB$ for $\mathcal{S}_N$ is an $\R^p_{\geqq}$-closed and $\R^p_{\geqq}$-bounded set such that $\mathcal{S}_N \subset \LB + \R^p_{\geqq}$ and $\LB = \LB_N$.
        \item An upper bound set $\UB$ for $\mathcal{S}_N$ is an $\R^p_{\geqq}$-closed and $\R^p_{\geqq}$-bounded set such that $\mathcal{S}_N \subset \mbox{cl}[\mathbb{R}^p \backslash (\mathcal{U} + \mathbb{R}_{\geqq}^p)]$ and $\mathcal{U} = \mathcal{U}_N $, where $\mbox{cl}(.)$ denotes the closure operator.
    \end{itemize}
    \label{def:boundSets}
\end{definition}

A particular lower bound set is the singleton $\{y^I\}$, where $y^I$, called the \emph{ideal point}, is defined by $y^I_k = \min_{y \in \YN}\{y_k\}$. Similarly, one can define the upper bound set $\{y^N\}$ where $y^N$, the \emph{nadir point}, is such that $y^N_k = \max_{y \in \YN}\{y_k\}$. %Beside, 

The linear relaxation of a MOILP $P$ is the problem $P^{LP} : \min \{z(x):x\in \mathcal{X}^{LP}\}$, where $\mathcal{X}^{LP} = \{x \in [0,1]: Ax \geqq b\}$. The problem $P^{LP}$ belongs to the class of \emph{Multi-Objective Continuous Linear Problem} (MOCLP). \cite{Ehrgott07} showed that solving the linear relaxation yields a valid lower bound set.

Given an upper bound set $\UB$, \cite{Klamroth15} proposed an alternative description of the region $\mbox{cl}[\mathbb{R}^p \backslash (\mathcal{U} + \mathbb{R}_{\geqq}^p)]$ using the \emph{set of local upper bounds} $\NU$. Let $u \in \Rp$, we define $C(u) = \{z \in \Rp : z \leqq u\}$. Using the definition of \cite{Klamroth15}, the set of local upper bounds $\NU$ is the set such that $\bigcup_{u \in \NU} C(u) = \mbox{cl}[\mathbb{R}^p \backslash (\mathcal{U} + \mathbb{R}_{\geqq}^p)]$, and for all $u^1, u^2 \in \NU$, $C(u^1) \nsubseteq C(u^2)$. The first condition makes sure that $\NU$ describes properly $\mbox{cl}[\mathbb{R}^p \backslash (\mathcal{U} + \mathbb{R}_{\geqq}^p)]$, whereas the second condition implies that there is no pair of local upper bounds such that one dominates the other, i.e., $\NU$ is of minimal size.%\comment[id=NF]{Figure for local upper bounds / search region? SP: NO I THINK IT IS OK LIKE THAT FIG 1 SHOWS IT THEN ANYWAY.}

Given a lower bound set $\LB$ and an upper bound set $\UB$, we define the search region as the set $\LB + \R^p_{\geqq} \cap \mbox{cl}[\mathbb{R}^p \backslash (\mathcal{U} + \mathbb{R}_{\geqq}^p)]$. The search region can be interpreted as the region of the objective space where non-dominated points are possibly located.

A \emph{weighted-sum scalarization} $P_{\lambda}$ of a MOILP $P$ is a single-objective optimization problem where the objective function is  %obtained by making 
a weighted sum of the objective functions of $P$. Hence, given a weight vector $\lambda \in \Rpgeq$, the problem $P_{\lambda}$ is written as $P_{\lambda} : \min\{\lambda z(x):x\in \mathcal{X}\}$.

\begin{property}{\citep{Ehrgott05}}
    \label{ppt:ws}
    Let $P$ be a MOCLP. Its non-dominated set $\YN$ corresponds to the non-dominated part of a polyhedron, and for any weighted-sum scalarization $P_{\lambda}$ with weight $\lambda \in \R_{\geqslant}$, there is an extreme point of $\YN$ that is optimal for $P_{\lambda}$.
\end{property}

Given a MOILP $P$, since its linear relaxation $P^{LP}$ is a MOCLP, Property~\ref{ppt:ws} implies that the optimal solution of a weighted-sum scalarization of $P^{LP}$ can be obtained by searching for the extreme point of the lower bound set that has the minimal weighted-sum of its objective values.
%looking NOT CLEAR WHAT "looking at" MEANS} at an extreme point $\LB$.
This property will be exploited in Section~\ref{sec:wsRule}.

\section{Related work}
\label{sec:relatedWorks}

To our knowledge, the first MOBB was proposed by \cite{KleinHannan1982}. In their algorithm, the authors use a single branching tree to solve a series of single-objective problems to generate all desired solutions. Later, \cite{Kiziltan1983} proposed a framework that uses the minimal completion, which consists of setting variables to 0 or 1 depending on their objective coefficients, %NOT CLEAR WHAT MINIMAL COMPLETION IS PLEASE EXPLAIN}
to generate lower bounds. The resulting solution is integer but is not necessarily feasible for the initial problem.

In the following decade, a lot of attention was paid to %the resolution of specific problems using a 
DSS approaches tailored to specific problems. We refer the reader to \cite{Ramos1998} and \cite{ViseeUlunguTeghem1998} for studies on the minimum spanning tree problem and the knapsack problem, respectively. In the latter, the novelty lies in the fact that they use a branch-and-bound algorithm in % to solve 
the second phase of a two-phase method, a well-known OSS algorithm proposed by \cite{UlunguTeghem1995}. In other words, they embedded a DSS algorithm %in
into an OSS algorithm%to solve the problem
, resulting in the first hybrid method.

%A straightforward lower bound in multi-objective optimization is provided by the ideal point.}
In multi-objective optimization, the ideal point provides a straightforward lower bound set.
The first to introduce more complex lower bounds in a DSS algorithm are \citet{SourdSpanjaard2008}. In their paper, the authors use a surface as a lower bound set, 
%EXPLAIN LOWER BOUND SET HERE OR MAYBE MOVE RELATED WORK TO AFTER DEFINITONS ETC.?}
namely the convex relaxation. Thereafter, the linear relaxation, weighted-sum scalarizations, and the linear relaxations of weighted sum scalarizations have been % also 
widely used in a similar way (see e.g., \cite{Vincent2013corrections}, \cite{StidsenAndersenDammann2014}, \cite{Belotti_2016}, \cite{Stidsen18}, \cite{parragh2019branch}, \cite{Gadegaard2019}, \cite{adelgren2021branch}). Although all these studies focus on the bi-objective case, the separating hypersurface principle from \cite{SourdSpanjaard2008} is also applicable in higher dimensions.
%Recently, \cite{DeSantis2020} and \cite{Forget2022} used hyperplanes and the linear relaxation respectively as lower bound sets for MOILP with three and more objective functions.
Recently, \cite{DeSantis2020} generated hyperplanes to obtain lower bound sets for multi-objective convex optimization problems (thus including MOILP) with three or more objective functions, whereas \cite{Forget2022} proposed to solve the linear relaxation for MOILP using more than two objectives. In the latter, the authors emphasized the difficulties raised by adding a third objective function. Indeed, if a simple dichotomic search is sufficient to calculate the linear relaxation with two objectives, things become more difficult when more dimensions are considered. In their paper, the authors suggest using Benson's outer approximation algorithm \citep{Benson1998,Hamel2013,BensolveA} to compute the linear relaxation based lower bound set.

\emph{Multi-Objective Mixed-Integer Linear Problems} (MOMILP), i.e., problems with both continuous and integer variables, have also received %also receive 
some attention in the MOBB literature. \citet{mavrotas1998branch} proposed a branch-and-bound framework that can handle MOMILP, as well as an improved version of their algorithm in \cite{mavrotas2005branch}. Later, \cite{Vincent2013corrections} proposed a refined version of their framework for the bi-objective 0-1 case. The use of MOBB to solve bi-objective MOMILP was further studied by \cite{Belotti_2016} and \cite{adelgren2021branch}.

In their paper, \cite{Vincent2013corrections} also conducted %proposed 
a study of different node selection rules. %In their experiments, 
They tested depth-first and breadth-first strategies on %SAY WHICH INSTANCES}
randomly generated instances, and depth-first appeared to be the most efficient. Similarly, \cite{parragh2019branch} tested both approaches on a different set of instances, but breadth-first performed the best. This suggests that the performance of the two classical node selection rules used in the literature, namely depth-first and breadth-first, are, in fact, dependent on the problem class. A similar observation was made in the preliminary study of \cite{Forget2022}, where both rules resulted in very different CPU times depending on the problem class of the instance solved. This issue is addressed in Sections~\ref{sec:nodeSel} and \ref{sec:expe}.
% \comment[id=NF]{Should we re-emphasize here that this is a gap that we address in this paper? YES!}

In the past decade, a lot of attention has been paid to methods that hybridize DSS and OSS algorithms. For the bi-objective case, \cite{StidsenAndersenDammann2014} proposed partitioning the objective space into multiple slices, leading to stronger upper bound sets. This also opened the door to parallelization, which was exploited in \citep{Stidsen18}, and resulted in promising improvements in performance. The authors also developed the concept of \emph{Pareto branching} (or \emph{objective (space) branching}): when the upper bound set partially dominates the lower bound set, it is possible to create disjoint sub-problems in the objective space by adding upper bounds on the objective functions to discard dominated regions from the search. This principle was further explored and improved independently by \cite{Gadegaard2019} and \cite{parragh2019branch}. In both papers, their experiments showed the great efficiency of this technique for the bi-objective case. Later, \cite{adelgren2021branch} also showed promising results using objective branching in MOBB applied to bi-objective MOMILP.

\cite{Forget21OB} extended objective branching to the multi-objective case. In their paper, the authors highlighted several challenges that arise when three or more objective functions are considered. In particular, they established that generating sub-problems without redundancies is a much more complex task compared to the bi-objective case, and they proposed a new method to overcome these difficulties. As a consequence, although still beneficial, objective branching did not appear to be as efficient as in the bi-objective case in their computational study. In this paper, we improve this result by showing that combining %probing with 
objective branching with probing results in a significant speed-up. %\comment[id=NF]{Should we re-emphasize here that this is a gap that we address in this paper?}
%Thus, just creating smaller disjoint sub-problems in the objective space is not necessarily enough to get significant speed-ups in terms of CPU time when three or more objective functions are considered.

Recently, \cite{adelgren2021branch} have proposed to use probing %, a popular technique used in the single-objective case, 
to enhance their bi-objective branch-and-bound framework. %The idea behind probing is to fix a variable to a particular value, and study the consequences of that action e.g. on the objective value, the feasibility of the problems, the values of the other variables, ect. 
The probing procedure of \cite{adelgren2021branch} relies on solving the 
%resolution of 
bi-objective linear relaxation based bound set, and showed promising results in their experiments. However, the impact of probing for problems with three or more objectives is unclear, as bound sets are %the linear relaxation is 
more complex to compute. Furthermore, objective branching cannot be applied as often and easily as in the bi-objective case, which may also have an impact on the performance of probing. % as well.
\section{Branch-and-bound framework}
\label{sec:BB}

The branch-and-bound framework developed %used 
in this paper is based on the framework of %from
\cite{Forget2022}, and is presented in this section. Similarly to the single-objective case, the principle is to divide a problem that is too hard to be solved into easier sub-problems. Each sub-problem is stored in a node, and the nodes are grouped together into a tree data structure $\nodeList$. For each node $\eta$, the sub-problem contained in $\eta$ is called $P(\eta)$, and each subproblem of $P(\eta)$ is stored in a child node of $\eta$. 
Instead of single numerical values, 
%The multi-objective nature of the problem implies that multiple solutions have to be reached in the tree, and thus
lower and upper bounds sets are used to determine whether a given sub-problem potentially %possibly 
contains new non-dominated solutions, which are feasible for the initial problem. If not, the corresponding node is fathomed. Otherwise, it is divided into disjoint sub-problems. A general outline of our framework is presented in Algorithm~\ref{alg:BB}.
%  that is not dominated by any the ones already known (stored in the upper bound set).

\begin{algorithm}[tb]
	\caption{Branch-and-bound algorithm for MOCOs}
	\label{alg:BB}
	\begin{algorithmic}[1]
		\STATE Create the root node $\eta^0$; set $\nodeList \leftarrow \{\eta^0\}$ and $\UB \leftarrow \emptyset$ \label{l:iniQ}
		\WHILE {$\nodeList\ne \emptyset$}
			\STATE Select a node $\eta$ from $\nodeList$; Set $\nodeList \leftarrow \nodeList \backslash \{\eta\}$ \label{l:ns}
	    	\STATE Compute a local lower bound set for $P(\eta)$ \label{l:lb}
    		\STATE If possible, update the upper bound set $\UB$ \label{l:ub}
			\IF {$\eta$ cannot be fathomed}
			    \STATE Split $P(\eta)$ into disjoint subproblems $P(\eta^1),...,P(\eta^h)$, and store each in a unique child node of $\eta$ and add them to $\nodeList$. \label{l:split}
			\ENDIF
		\ENDWHILE
		\RETURN {$\UB$}
	\end{algorithmic}
\end{algorithm}

The branch-and-bound is initialized with an empty upper bound set $\UB$, and a list of non-explored nodes, denoted by $\nodeList$, that contains the initial problem (at the root node of the tree). %The list of non-explored nodes is denoted by $\nodeList$. 
At each iteration, a node $\eta$ is selected and removed from $\nodeList$ (Line~\ref{l:ns} of Algorithm~\ref{alg:BB}). The classical tree exploration strategies %methodologies
from the literature are depth-first (last in first out) and breadth-first (first in first out), and are further discussed in Section~\ref{sec:expe}. If there is no non-explored node remaining, the algorithm stops and $\UB = \YN$.
%YOU RETURN UB, SO IT SHOULD BE DIE OTHER WAY AROUND ($\UB = \YN$), NO?}

When a node $\eta$ is selected, a lower bound set $\LB(\eta)$ for $P(\eta)$ is computed (Line~4 of Algorithm~\ref{alg:BB}). In this framework, the linear relaxation $P^{LP}(\eta)$
%MAYBE BETTER $P^{LP}(\eta)$, NO?}
is solved, %at each node, 
and the result yields a valid lower bound set \citep{Ehrgott07}. A Benson-type algorithm is used for this purpose (see, e.g., \cite{Hamel2013}). As a result, a description of $\LB(\eta)$ in terms of its extreme points is obtained, as well as a description of $\LB(\eta) + \Rpgeq$ in terms of its hyperplanes.
%  : \min\{Cx : \X^{LP}(\eta)\}$, where $\X^{LP} = \{x\in[0,1]^n : A^{\eta}x \geqq b^{\eta}\}
% Note that $A^{\eta}$ and $b^{\eta}$ include both the initial constraints and all the branching constraints at node $\eta$. Here, the warm-starting procedure from \cite{Forget2022}, which rely on a Benson-type algorithm \citep{Benson1998, Hamel2013, BensolveA}, is used to solve it.

Once the lower bound set is obtained, if possible, new non-dominated points 
%I WOULD USE PRESENT TENSE EVERYWEHRE HERE} %will be
are
harvested (Line~5 of Algorithm~\ref{alg:BB}). Indeed, Benson's algorithm returns %after computation, 
a pre-image %is known 
for each extreme point of $\LB(\eta)$. Hence, any extreme point $l$ with an integer pre-image that is not dominated by any existing point in the upper bound set will be added to $\UB$; and all points $y \in \UB$ that are dominated by $l$ %such that 
($l \geq y$) are %will be 
removed from $\UB$.

We distinguish three cases in which a node can be fathomed:

%Note that 
(i) If $P^{LP}(\eta)$ is infeasible, no new non-dominated points are searched (i.e., Line~5 is skipped), and the node is \emph{fathomed by infeasibility}. Indeed, similarly to the single-objective case, if $P^{LP}(\eta)$ is infeasible, $P(\eta)$ is also infeasible. 

{(ii) I}f $\LB(\eta)$ is made of a unique extreme point $l$ with an integer pre-image, all new points found in $P(\eta)$ will be dominated by the integer solution $l$ and consequently, $\eta$ % the node 
is \emph{fathomed by optimality}.

(iii) Finally, a third way to fathom a node exists: \emph{fathoming by dominance}. This case happens when the lower bound set $\LB(\eta)$ is dominated by the upper bound set $\UB$. From the definitions, this situation is equivalent to saying that each feasible point of $P^{LP}(\eta)$, and thus of $P(\eta)$, is dominated by at least one already known integer point $u \in \UB$. Consequently, no new non-dominated point can be found in $P(\eta)$. In practice, if there exists no local upper bound $u \in \NU$ such that $u \in \LB(\eta) + \Rpgeq$, then the node is fathomed by dominance. This dominance test was first introduced by \cite{SourdSpanjaard2008}, and used multiple times in the literature (see, e.g., \cite{Gadegaard2019, Forget2022}).

If the node $\eta$ cannot be fathomed, we resort to \emph{branching}, and $P(\eta)$ is split into several sub-problems (Line~\ref{l:split} of Algorithm~\ref{alg:BB}). To do so, objective branching is used first. This technique was initially %first 
introduced for the bi-objective case by \cite{StidsenAndersenDammann2014}, further improved by \cite{Gadegaard2019} and \cite{parragh2019branch}, and finally extended to the multi-objective case in \cite{Forget21OB}. It consists of creating disjoint sub-problems in the objective space when the lower bound set is partially dominated by the upper bound set, with the purpose of discarding regions that cannot contain any   %where it is known that no 
new non-dominated points. % can be found. 
The subproblems are created by adding constraints in the form $z(x) \leqq s$, $s \in \Rp$, and the point $s$ is called \emph{super local upper bound}. Objective branching will be further elaborated upon in Section~\ref{sec:OB}.

Once objective branching is applied, a set of subproblems $\eta^1,...,\eta^{\gamma}$ is obtained. Note that only one subproblem is obtained ($\gamma = 1$) if it is not possible to create two or more disjoint subproblems in the objective space. Then, for each of the $\gamma$ subproblems, \emph{decision space branching} is performed. To do so, one variable $x_i$ is chosen, and two subproblems with the constraints $x_i = 0$ or $x_i = 1$ are %will be 
created. The variable $x_i$ has to be a free variable, i.e., not fixed to a specific value in the current node by one of the previous
branching decisions. %decision branching constraints.

\section{Objective branching induced enhancements}
%Enhancing} %Improving 
%objective branching}
\label{sec:OB}

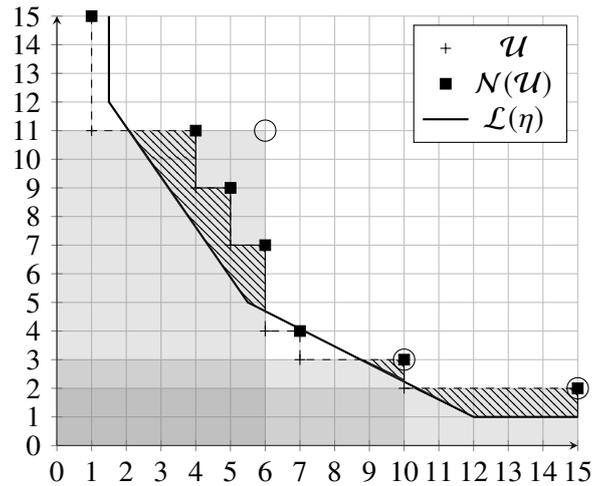
\begin{figure}
    \centering
    \begin{center}
\begin{tikzpicture}
            \begin{axis}[axis x line=bottom, axis y line = left, grid=major,
            xmin = 0, xmax = 15,
            ymin = 0, ymax = 15 , ytick={0,1,2,...,15} , xtick = {0,1,2,...,15}]

            % UB
            \addplot[only marks, mark=+] coordinates {(1,11) (4,9) (5,7) (6,4) (7,3) (10,2)};
            \addlegendentry{$\UB$}
            
            % Lubs
            \addplot[only marks,mark=square*] coordinates {(1,15) (4,11) (5,9) (6,7) (7,4) (10,3) (15,2)};
            \addlegendentry{$\NU$}

            % LB
            \addplot[thick] coordinates {(1.5,15) (1.5,12) (5.5,5) (12,1) (15,1)};
            \addlegendentry{$\LB(\eta)$}

            % cl(UB)
            \addplot[dashed, samples=100] coordinates{(1,15)(1,11)(4,11)(4,9)(5,9)(5,7)(6,7)(6,4)(7,4)(7,3)(10,3)(10,2)(15,2)};

            % OB subpb
            \path[fill = gray, opacity = 0.2] (0,0) -- (6,0) -- (6,11) -- (0,11);
            \path[fill = gray, opacity = 0.2] (0,0) -- (10,0) -- (10,3) -- (0,3);
            \path[fill = gray, opacity = 0.2] (0,0) -- (15,0) -- (15,2) -- (0,2);
            %\addlegendentry{subproblems}
            
            % Slubs
            \addplot[only marks,mark=o,mark size=4pt] coordinates {(6,11) (10,3) (15,2)};
            
            % Search region
            \addplot[pattern=north west lines] coordinates {(2.1,11) (5.5,5) (6,4.7) (6,7) (5,7) (5,9) (4,9) (4,11)};
            \addplot[pattern=north west lines] coordinates {(8.8,3) (10,2.3) (10,3)};
            \addplot[pattern=north west lines] coordinates {(10.3,2) (12,1) (15,1) (15,2)};

            \end{axis}
            \end{tikzpicture}
                
\end{center}
    \caption{The lower bound set $\LB(\eta)$, depicted by the solid line, is partially dominated by the upper bound set $\UB$, represented by the crosses. In this situation, it is possible to split the problem into three disjoint sub-problems in the objective space. Each sub-problem is highlighted by the hatched areas, and its corresponding super local upper bound is depicted by its closest large circle.}
    \label{fig:OB2obj}
\end{figure}

%MAYBE A BETTER/MORE CATCHY SENTENCE CAN BE FOUND INSTEAD OF "In this section..."?}
%In this section, we further 
We now elaborate on %explore 
the concept of objective branching and how it can be exploited to enhance our MOBB. %In
Figure~\ref{fig:OB2obj} %is 
depicts a situation where the lower bound set $\LB(\eta)$ is partially dominated by the upper bound set $\UB$, and the resulting search region is given by the hashed areas. Any part of the objective space that is not included in one of these areas cannot contain any feasible non-dominated point for $P(\eta)$. Objective branching consists in generating disjoint subproblems in a way such that as much of the region of the objective space dominated by $\UB$ is discarded from all subproblems, without excluding any point of the search region from the sub-problems. In the example from Figure~\ref{fig:OB2obj}, this results in three subproblems, defined by three super local upper bounds depicted by the large circles.% Also, note that every point of the objective space that should be included in at least one of the subproblems (with respect to the objective branching constaints only) is located in the gray area, otherwise we may miss a non-dominated point.

In the bi-objective case, multiple ways to compute the subproblems exist. \cite{StidsenAndersenDammann2014} and \cite{Gadegaard2019} generated new subproblems when the algorithm detected that one or several points of the upper bound set partially dominate the lower bound set, whereas \cite{parragh2019branch} kept track of the various non-dominated segments of the lower bound set and generated a subproblem for each. The two approaches are equivalent in the sense that exactly the same subproblems are generated with both methods.

Recently, \cite{Forget21OB} showed that the computation of the subproblems in the case where $p \geq 3$ was more complex but still possible. However, the increased complexity resulted in a less significant benefit of using objective branching when $p \geq 3$ compared to the case where $p = 2$. For some problem classes, it even resulted in worse computation times, which is in % creates a 
great contrast to %with 
the bi-objective case, where using objective branching systematically led to lower CPU times. As a result, it appears that when $p \geq 3$, the use of objective branching is not always sufficient by itself, and in this paper, we aim to study whether objective branching constraints can be further exploited %utilized 
to help reduce %ing 
the total CPU time of the branch-and-bound framework. %\comment[id=NF]{Should this paragraph go to the introduction?}

    % ---------------------------------------
    %   Variable fixing
    % ---------------------------------------

    \subsection{Probing} 
    %{Variable fixing}
    \label{sec:varFix}

\begin{figure}
        \centering
        \begin{minipage}{0.5\textwidth}
            \centering
            \includegraphics[width=\textwidth]{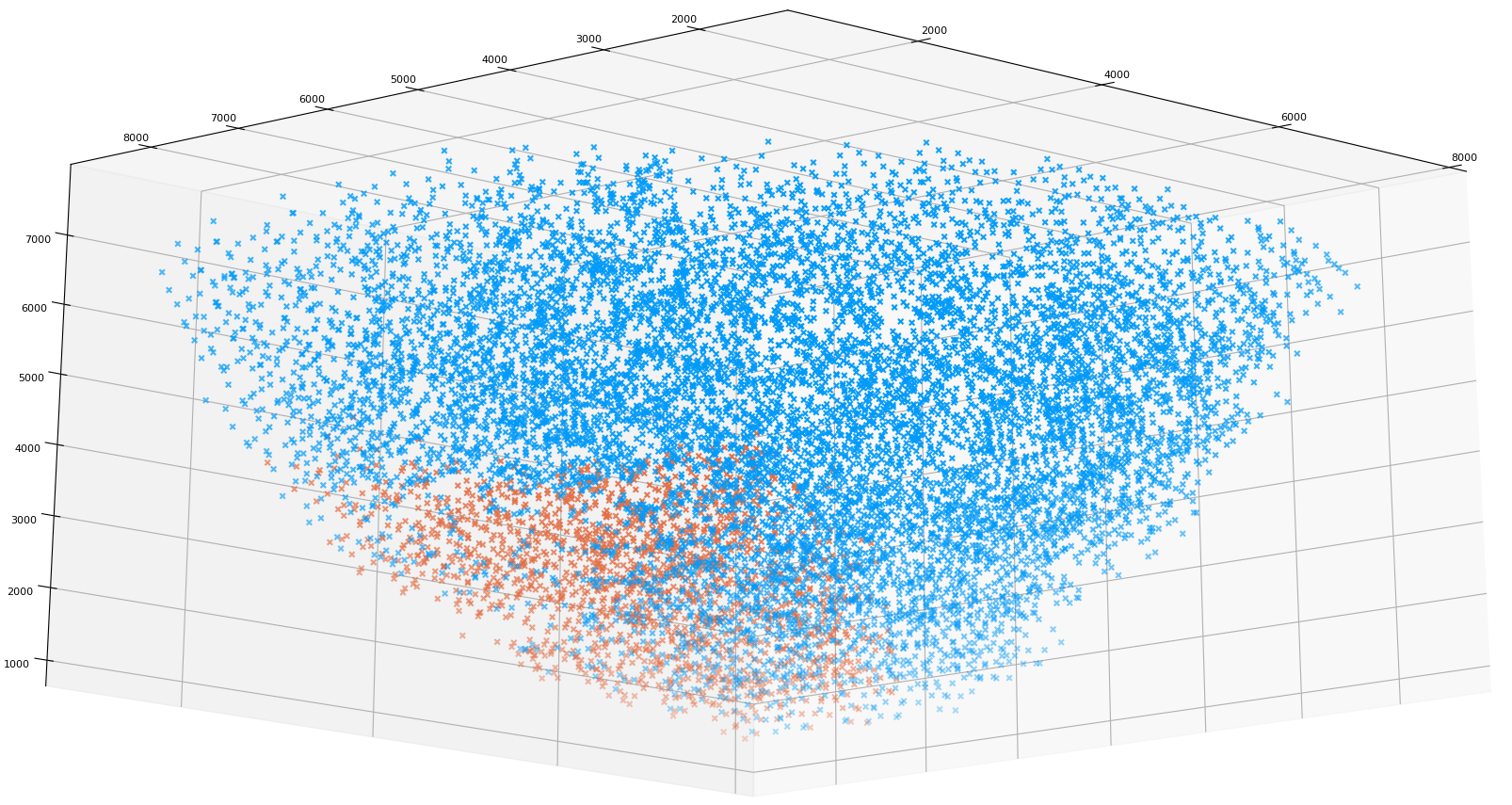} % first figure itself
            %\caption{Uncapacited Facility Location Problem with 4672 non-dominated points.}
        \end{minipage}\hfill
        \begin{minipage}{0.5\textwidth}
            \centering
            \includegraphics[width=\textwidth]{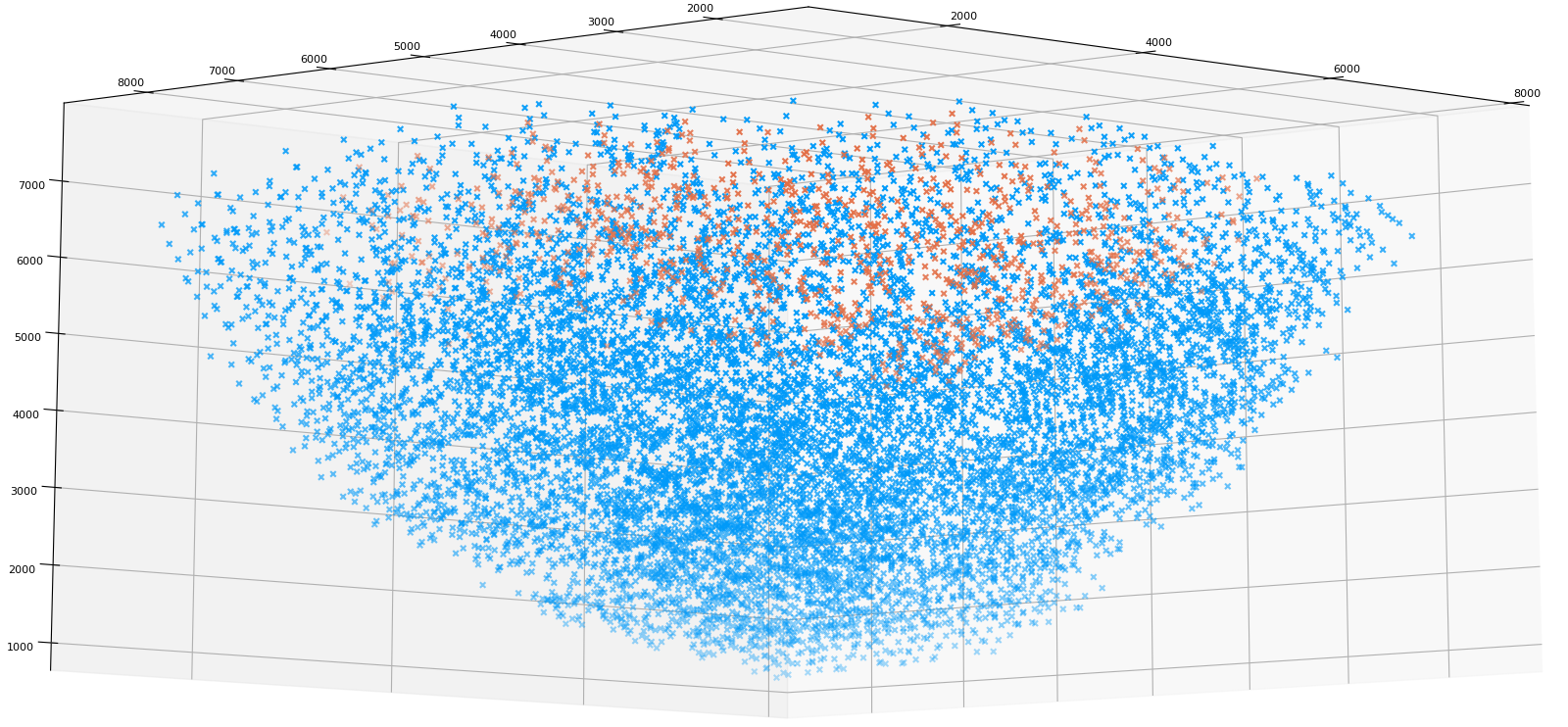} % second figure itself
            %\caption{Uncapacited Facility Location Problem with 4672 non-dominated points.}
        \end{minipage}
    
        \begin{minipage}{0.5\textwidth}
            \centering
            \includegraphics[width=\textwidth]{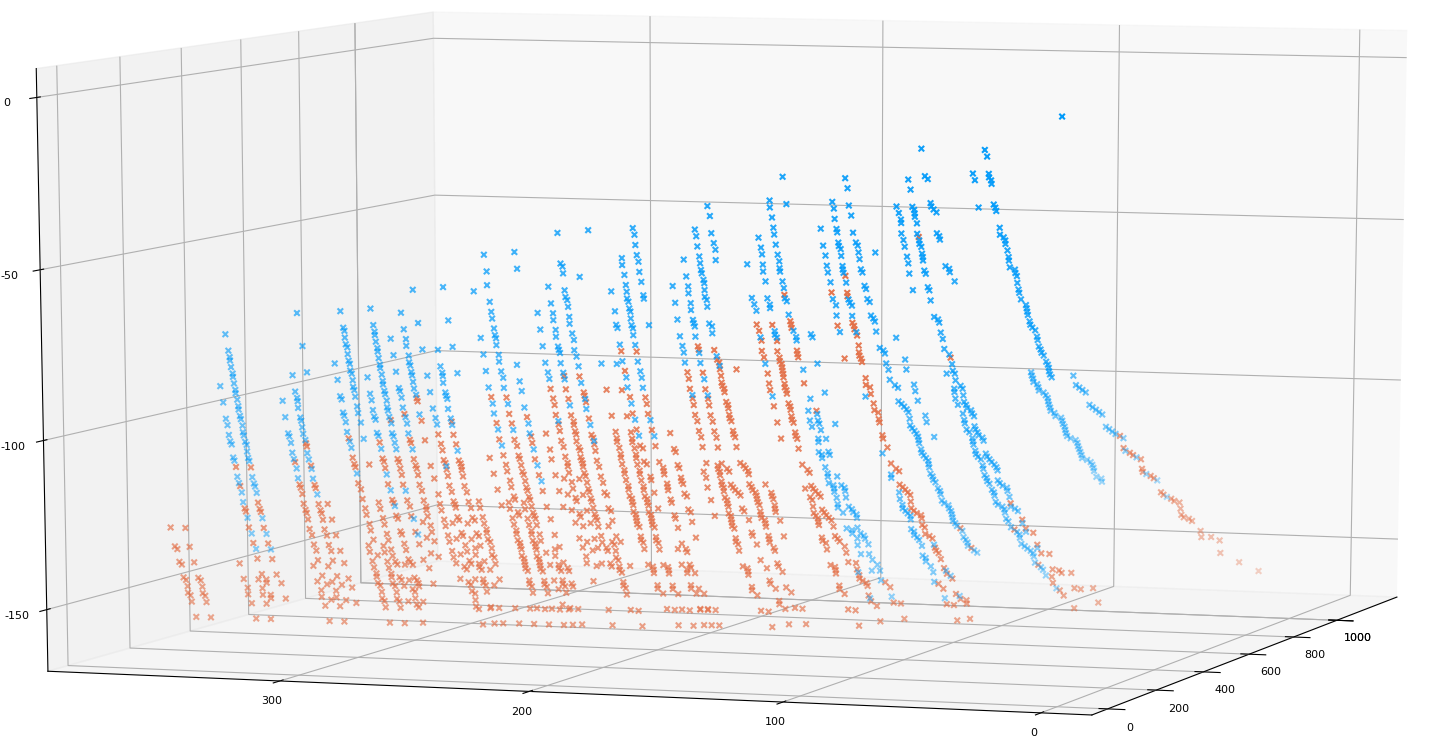} % first figure itself
            %\caption{Capacited Facility Location Problem with 4672 non-dominated points.}
        \end{minipage}\hfill
        \begin{minipage}{0.5\textwidth}
            \centering
            \includegraphics[width=\textwidth]{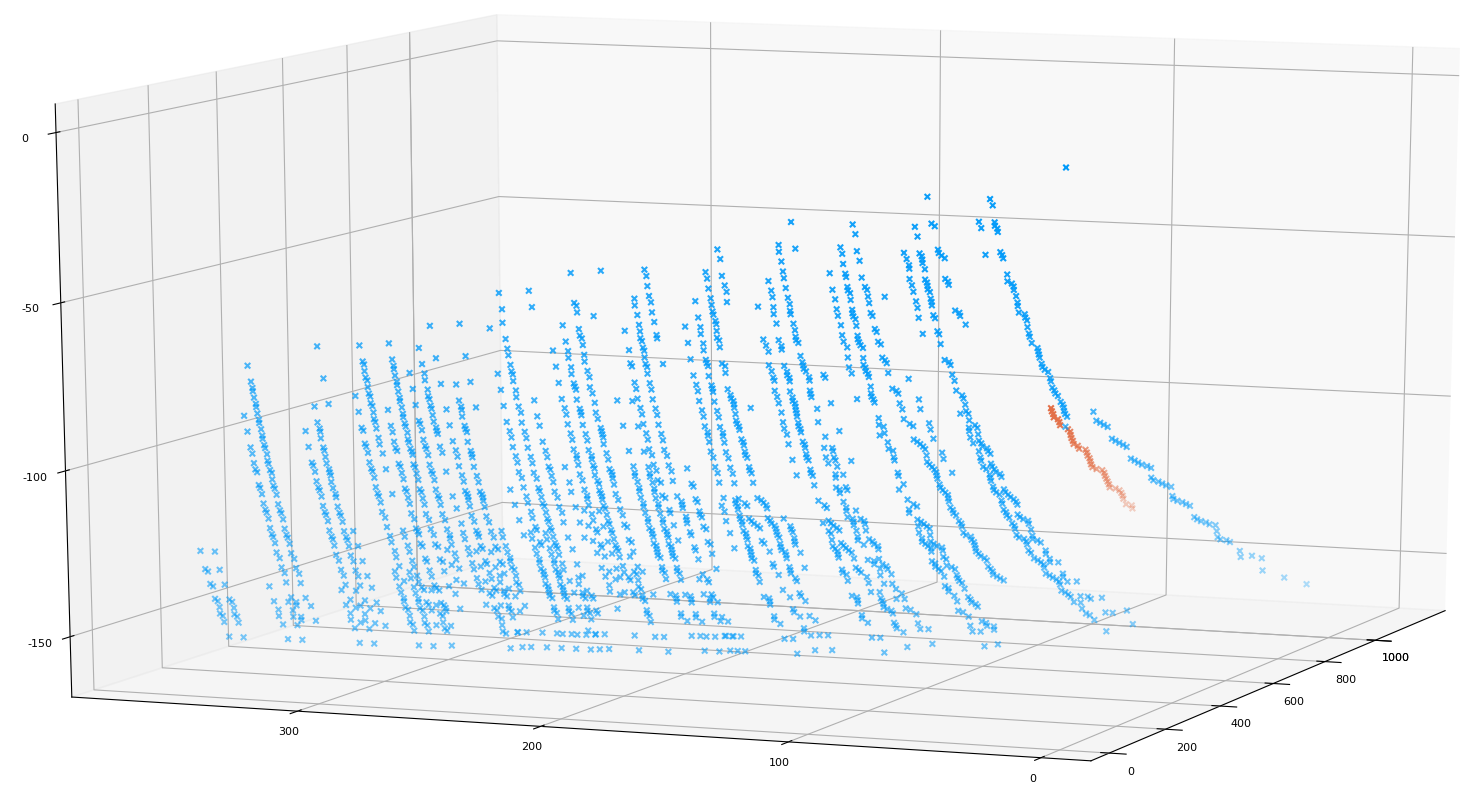} % second figure itself
            %\caption{Capacited Facility Location Problem with 4672 non-dominated points.}
        \end{minipage}

        \caption{The first row depicts the set of non-dominated points for a tri-objective Uncapacitated Facility Location Problem with 4672 non-dominated points. The second row depicts the set of non-dominated points for a tri-objective Capacitated Facility Location Problem with 1912 non-dominated points. For each plot, a variable $x_i$ was chosen. A blue point is a non-dominated point where $x_i = 0$ whereas an orange point is a non-dominated point for which $x_i = 1$.}
        \label{fig:valVar}
    \end{figure}

    % To do so, we rely on the observation that t
    %wo points that are close to each other in the objective space are likely to have similar pre-images.
    In multi-objective optimization, an intuitive belief is that two points that are close to each other in the objective space are more likely to have similar pre-images than two points that are far away from each other.
    %In
    Figure~\ref{fig:valVar} shows %is plotted 
    the set of non-dominated points of %for 
    two %instances of 
    tri-objective MOCO instances (one row for each instance). A blue point corresponds to a non-dominated point where the chosen variable $x_i$ takes value $0$, whereas an orange point corresponds to a non-dominated point such that $x_i = 1$. From these four pictures, it is clear that some problem classes have variables that take a particular value in certain parts of the objective space.
    Moreover, when applying objective branching, the algorithm reduces the search to particular regions of the objective space. Hence, based on the previous observation, it is possible that some variables cannot take specific values %anymore 
    in certain subproblems. %Here,
    The process of identifying such values is commonly referred to as probing. In the following, we present our probing strategies. %We aim to identify such variables.
    %  for points

    %\comment[id=NF]{make black and white plots in Figure~\ref{fig:valVar}?}
    
    % ----------------------------------------
    % naive strat
    
    \subsubsection{A naive strategy}
    
    We explore at first a naive strategy. At node $\eta$, we define the set of variables fixed to $0$ and $1$ as $\mathcal{I}^0(\eta) = \{i \in \{1,...,n\} : x_i = 0\}$ and $\mathcal{I}^1(\eta) = \{i \in \{1,...,n\} : x_i = 1\}$ respectively. The set of free variables is $\mathcal{I}^f(\eta) = \{1,...,n\} \backslash (\mathcal{I}^0(\eta) \cup \mathcal{I}^1(\eta)\}$. With a little abuse of notation, %By abuse of language, 
    we will consider that writing $x_i \in \mathcal{I}^0(\eta)$ is equivalent to $i \in \mathcal{I}^0(\eta)$. We will consider analogous statements for $\mathcal{I}^1(\eta)$ and $\mathcal{I}^f(\eta)$. % as well.
    
    Let $x_i \in \mathcal{I}^f(\eta)$ be a free variable at node $\eta$. Since we consider problems with binary variables only, the possible values for $x_i$ are $0$ or $1$. A first approach in order to check whether $x_i$ can take value $v \in \{0,1\}$ in $P(\eta)$ is to solve the linear program $F(i,v) : \min\{0 \ | \ x \in \X^{LP}(\eta), \ x_i = v\}$. If $F(i,v)$ is not feasible, then $x_i$ cannot take value $v$. When both $F(i,0)$ and $F(i,1)$ are solved for $x_i$, there are four possible scenarios:%four possible scenarios are identified.
    
    \begin{itemize}
        \item Both $F(i,0)$ and $F(i,1)$ are feasible: nothing can be concluded about $x_i$, and thus, the variable remains a free variable;
        \item $F(i,0)$ is feasible and $F(i,1)$ is infeasible: $x_i$ is fixed to $0$;%, i.e. $\mathcal{I}^f(\eta) \leftarrow \mathcal{I}^f(\eta) \backslash \{i\}$ and $\mathcal{I}^0(\eta) \leftarrow \mathcal{I}^0(\eta) \cup \{i\}$;
        \item $F(i,0)$ is infeasible and $F(i,1)$ is feasible: $x_i$ is fixed to $1$;%, i.e. $\mathcal{I}^f(\eta) \leftarrow \mathcal{I}^f(\eta) \backslash \{i\}$ and $\mathcal{I}^1(\eta) \leftarrow \mathcal{I}^1(\eta) \cup \{i\}$;
        \item Both $F(i,0)$ and $F(i,1)$ are infeasible: there is no possible integer value for $x_i$. Thus, the node $\eta$ is fathomed by infeasibility.
    \end{itemize}
    
    When $x_i$ is fixed to $0$, the set of free variables $\mathcal{I}^f(\eta)$ is updated to $\mathcal{I}^f(\eta) \backslash \{i\}$ since $x_i$ is not free anymore, and $\mathcal{I}^0(\eta)$ becomes $\mathcal{I}^0(\eta) \cup \{i\}$. Similarly, if $x_i$ is fixed to $1$, then $\mathcal{I}^f(\eta)$ and $\mathcal{I}^1(\eta)$ become $\mathcal{I}^f(\eta) \backslash \{i\}$ and $\mathcal{I}^1(\eta) \cup \{i\}$ respectively. If all free variables are fixed to a particular value, the node is fathomed by optimality. Indeed, this situation implies that there is only one integer solution in $P(\eta)$, and no new non-dominated point can be reached in $P(\eta)$. The upper bound set is updated with the new point obtained by fixing all variables.
    
    A first naive strategy is to solve $F(i,v)$ at each node $\eta$, for each free variable $x_i \in \mathcal{I}^f(\eta)$, and for each possible value $v \in \{0,1\}$. The approach is similar to \cite{adelgren2021branch} in the sense that they also perform probing at each node. In their paper, the authors suggest to perform probing both before the computation of the linear relaxation, and when creating sub-problems. In the latter case, they apply probing after selecting a free variable to branch on, and change the branching variable if they conclude that the decision led to an infeasible problem. In this paper, we adopted a slightly different approach: we perform probing only after objective branching and before variable branching. % only. 
    In this way, %That way, 
    we aim to reduce the set of branching candidates at each node, while still benefiting from the objective branching constraints. Indeed, we expect these to be the most constraining to the problem, since they restrict the search to a particular region of the objective space and thus, hopefully, reduce the possible values taken by the variables and help the algorithm to make an appropriate branching decision.
    
    Another difference to %with
    \cite{adelgren2021branch} is that when performing probing, they solve the bi-objective linear relaxation of the corresponding problem instead of solving a simple feasibility problem as we do. However, as we consider more objective functions, the linear relaxation becomes significantly more expensive to compute. 
    %I WOULD REMOVE THIS SENTENCE:"This implies that a huge effort in terms of CPU time will be required in order to compute multiple times the linear relaxation, even partially."}
    %and it quickly becomes a huge burden in terms of cpu time to re-compute multiple times even partially the linear relaxation.
    Consequently, our approach requires at most one single-objective linear program to be solved for each variable and value, which, in the binary case, limits the maximum number of LPs to be solved to $2 |\mathcal{I}^f(\eta)|$ at each node. %amount of effort put into the procedure.
    % Hence, when fixing variables, we opted for weaker but less expensive lower bound sets, namely the linear relaxation of a weighted sum scalarization. As presented in the Section~\ref{sec:varFix}, t
    
    % ---------------------------
    % Saving LPs
    
    \subsubsection{An advanced strategy}
    \label{sec:vfAdvStrat}
    
   % In this section, we improve the naive strategy by reducing the number of linear programs to solve in order to fix variables. 
   The naive strategy can be improved.
   Indeed, it is possible in some cases to detect if $F(i,v)$ is feasible without actually solving the linear program. For instance, if a solution that is feasible for $F(i,v)$ is already known, there is no need to solve $F(i,v)$. Such solutions can be collected, for example, from the extreme points of the lower bound set, or by keeping track of the solutions obtained from previously solved linear programs ($F(j,v')$, $j \ne i$, $v' \in \{0,1\}$) in the current node. 
    
    Moreover, variables can be fixed by inspection. % one can try to fix variables by looking manually at the constraints.
    %If in constraint $i$, we have $a_{ij}x_j \leq b_i$ where $a_{ij} > b_i$, then $x_j$ cannot take value $1$. Otherwise, the constraint would be violated. Such constraint does not appear in the initial problem (assuming that it is feasible), but may happen deeper in the tree as multiple variables are fixed through variable space branching constraints and variable fixing, reducing the number of actual variables in the constraints.
    Let $x_j \in \mathcal{I}^f(\eta)$ be a free variable, and $\sum_{l=1}^n a_{il}x_l \leq b_i$ a constraint of the problem. By comparing $a_{ij}$ to the maximal possible value of the right-hand side of the constraint, one may be able to conclude that $x_i$ cannot take value $1$.
    First, all variables fixed to $1$ are considered as constants and will be used to adjust the right-hand side. Then, all free variables $x_l \in \mathcal{I}^f(\eta)$ such that $a_{il} \leq 0$ will be temporarily fixed to $0$ meanwhile, those where $a_{il} < 0$ will be temporarily fixed to $1$. The constraint then becomes $a_{ij}x_j \leq b - \sum_{l \in \mathcal{I}^1(\eta)} a_{il} + \sum_{l \in \mathcal{I}^f(\eta), a_{il} < 0, l \ne j} a_{il}$. Then, if $a_{ij} > b - \sum_{l \in \mathcal{I}^1(\eta)} a_{il} - \sum_{l \in \mathcal{I}^f(\eta), a_{il} < 0, l \ne j} a_{il}$, the variable $x_j$ can be fixed to $0$. Similar rules can be used for constraints in the form $\sum_{l=1}^n a_{il}x_l \geq b_i$ and $\sum_{l=1}^n -a_{il}x_l \leq -b_i$. %Equality constraints ($\sum_{l=1}^n a_{il}x_l = b_i$) can be checked for example by applying the same rule to both $\sum_{l=1}^n a_{il}x_l \leq b_i$ and $\sum_{l=1}^n a_{il}x_l \geq b_i$.
    More complex %preprocessing 
    rules could be used as well, but this is out-of-scope  %falls out of the scope 
    of this paper and thus, we will stick to these simple rules here.
    % The same rule can be applied to any
    % by multiplying the constraint by $-1$, which results in a constraint in the form
    
    We note that probing may be applied after variable branching as well as after objective branching. However, our experiments showed that it is most effective when used in conjunction with objective branching. A possible explanation relates to the example given in Figure~\ref{fig:valVar}: some variables may only take certain values in certain regions of the objective space and objective branching induces such regions.
    %Finally, preliminary experiments and results in Section~\ref{sec:expe} showed evidence that probing is the most effective when coupled with objective branching. This is due to the fact that objective branching constraints ???
    %this is used in combination with objective branching that probing led to the most important speed-up.
    Hence, at node $\eta$, we propose to apply probing only if an improvement in the objective branching constraints is observed compared to its parent %his father 
    node $\hat{\eta}$. In other words, we perform %compute 
    probing only if $s \leqslant \hat{s}$, where $s$ and $\hat{s}$ are the super local upper bounds defining the objective branching constraints in nodes $\eta$ and $\hat{\eta}$, respectively.
    
    This advanced strategy is new compared to \cite{adelgren2021branch}, as we first aim at achieving the same results by solving less linear programs, and then suggest using probing only when it is expected to be the most relevant.
    
    % -------------------------------
    % using the obj function
    
    \subsubsection{Combining probing and bounding}
    %{Using the objective function}
    \label{sec:objFunc}
    
    %Finally, 
    The linear program $F(i,v)$ solved when applying the naive strategy does not have an objective function.
    %However, by introducing a weighted sum of the objectives %\comment[id=NF]{define ws-scalar in Section~\ref{sec:def}?}
    %as the objective function, it is possible to obtain additional information. 
    However, using an objective function may provide us with additional information. For this purpose, we rely on adding a weighted sum objective function.
    Given a weight vector $\lambda \in \Rp$, the linear program $F(i,v,\lambda) : \min\{\lambda z(x) \ | \ x \in \X^{LP}(\eta), \ x_i = v\}$ is solved, and the optimal value $z^{*v}$ is obtained. By definition, when $x_i$ is fixed to $v$ in this sub-problem, all feasible solutions $x \in \X(\eta) \cap \{x_i = v\}$ are such that $\lambda z(x) \geq z^{*v}$. Moreover, because of the objective branching constraints, all feasible solutions $x \in \X(\eta) \cap \{x_i = v\}$ are such that $z(x) \leqq s$, $s \in \Rp$. Hence, we can conclude that if there is no local upper bound $u \in \NU$ such that $\lambda u \geq z^{*v}$ and $u \leqq s$, then there is no feasible solution $x \in \X(\eta) \cap \{x_i = v\}$ that can generate a new non-dominated point. In this case, $x_i$ is fixed to $1$ if $v = 0$, or to $0$ if $v = 1$. Note that the dominance test we employ here has, e.g., also bee used by %this is very similar to the dominance test used in 
    \cite{StidsenAndersenDammann2014,Stidsen18} in a bi-objective context.
    
    An example is given in Figure~\ref{fig:varFixDomi}, where $\lambda = (1,1)$ is used, and the programs $F(i, 0, \lambda)$ and $F(i, 1, \lambda)$ are solved. In the rightmost figure, $F(i, 0, \lambda)$ is feasible, and the weighted-sum resulted in a non-empty search region. On the contrary, in the leftmost figure, $F(i, 1, \lambda)$ is feasible but the weighted-sum is dominated by the upper bound set in the region considered. This implies that in this sub-problem, all integer solutions in which $x_i$ takes value $1$ are dominated by at least one existing integer solution. Hence, there is no need to branch on $x_i$, and the variable can be fixed to $0$.
    % , with a reasoning similar to the dominance test mentioned in Section~\ref{sec:BB}
    
    %\begin{figure}
    %    \centering
    %    \begin{minipage}{width=0.5\textwidth}
    %        \input{tikzFigures/VarFixDomi.tex}
    %    \end{minipage}
    %    \begin{minipage}{width=0.5\textwidth}
    %        \input{tikzFigures/VarFixDomi.tex}
    %    \end{minipage}
    %    \caption{Caption}
    %    \label{fig:my_label}
    %\end{figure}
    
    \begin{figure}
        %\centering
        \begin{minipage}{.47\textwidth}
            \subcaptionbox{
            \small{Fixing $x_i=0$: the weighted-sum based LB set defined by $F(i,0,\lambda)$ (dash-dotted line) and the objective branching constraints (OB) is only partially dominated by the current UB set $\UB$.}}
            %\textbf{\small{The linear program $F(i,0,\lambda)$ was solved, and the optimal value $z^{*0}$ was obtained. All points of the objective space satisfying the objective branching constraints and $\lambda z(x) = z^{*0}$ are depicted by the dash-dotted line (weighted sum value). All feasible points are located both in the gray area (objective branching constraints) and above the dashdotted line. Since some local upper bounds are located in this region, $x_i = 0$ is a possible candidate for new non-dominated points in this region of the objective space.}}}
            %\label{fig:varFixDomi2}
            {%\begin{center}
    \begin{tikzpicture}[scale = 0.9]
        \begin{axis}[axis x line=bottom, axis y line = left, grid=major,
        xmin = 0, xmax = 24,
        ymin = 0, ymax = 16 , ytick={0,2,...,16} , xtick = {0,2,...,24}, xlabel = $z_1(x)$, ylabel = $z_2(x)$]

        % LUB
        \addplot[only marks,mark=square*] coordinates {(2,16) (6,14) (8,10) (10,8) (16,2) (24,1)};
        
        % Pts
        \addplot[only marks,mark = +] coordinates {(2,14) (6,10) (8,8) (10,2) (16,1)};
        % OB
        \addplot[only marks,mark=o,mark size=4pt] coordinates{(16,14)};
        
        % x_i = 0
        \addplot[thick, dashdotted] coordinates{(3,14)(16,1)};
        
        % x_i = 1
        %\addplot[thick, dashdotted] coordinates{(8,14)(16,6)};

        % LB
        %\addplot[thick, samples=100, domain=6:12, green,name path = A] {9 - 0.5*x};
        %\addplot[thick, samples=100, domain=4.5:6,green,name path = C] {18 - 2*x};

        % UB
        \addplot[dashed, samples=100] coordinates{(2,16)(2,14)(6,14)(6,10)(8,10)(8,8)(10,8)(10,2)(16,2)(16,1)(24,1)};

        % OB fill
        \path[fill = gray,opacity = 0.2] (0,14) -- (16,14) -- (16,0) -- (0,0);
        
        % Search region
        \addplot[pattern=north west lines] coordinates {(3,14) (6,11) (6,14)};
        \addplot[pattern=north west lines] coordinates {(7,10) (8,9) (8,10)};
        \addplot[pattern=north west lines] coordinates {(9,8) (10,7) (10,8)};
        \addplot[pattern=north west lines] coordinates {(15,2) (16,1) (16,2)};
        
        % UB fill
        %\path[fill = red,opacity = 0.1] (0,100) -- (10,100) -- (10,90) -- (30,90) -- (30,60) -- (50,60) -- (50,50) -- (70,50) -- (70,30) -- (80,30) -- (80,24) -- (90,24) -- (90,10) -- (130,10) -- (130,0) -- (0,0);

        \addlegendentry{$\NU$}
        \addlegendentry{$\UB$}
        \addlegendentry{OB}
        \addlegendentry{$x_i = 0$}
        %\addlegendentry{$x_i = 1$}
    %\addlegendentry{OB}

    \end{axis}
    \end{tikzpicture}
%\end{center}
}
            %\input{tikzFigures/VarFixDomi0}
            %\caption{The linear program $F(i,0,\lambda)$ was solved, and the optimal value $z^{*0}$ was obtained. All points of the objective space satisfying the objective branching constraints and $\lambda z(x) = z^{*0}$ are depicted by the dashdotted line (weighted sum value). All feasible points are located both in the gray area (objective branching constraints) and above the dashdotted line. Since some local upper bounds are located in this region, $x_i = 0$ is a possible candidate for new non-dominated points in this region of the objective space.}
        \end{minipage}
        \begin{minipage}{.06\textwidth}
            \ 
        \end{minipage}
        \begin{minipage}{.47\textwidth}
            \subcaptionbox{
             \small{Fixing $x_i=1$: the weighted-sum based LB set defined by $F(i,1,\lambda)$ (dash-dotted line) and the objective branching constraints (OB) is dominated by the current UB set $\UB$.}}
            %\textbf{\small{The linear program $F(i,1,\lambda)$ was solved, and the optimal value $z^{*1}$ was obtained. All points of the objective space satisfying the objective branching constraints and $\lambda z(x) = z^{*1}$ are depicted by the dash-dotted line. All feasible points are located both in the gray area (objective branching constraints) and above the dashdotted line (weighted sum value). Since no local upper bounds are located in this region, $x_i = 1$ is not a possible value for new non-dominated points in this region of the objective space.}}}
            %\label{fig:varFixDomi2}
            {%\begin{center}
    \begin{tikzpicture}[scale = 0.9]
        \begin{axis}[axis x line=bottom, axis y line = left, grid=major,
        xmin = 0, xmax = 24,
        ymin = 0, ymax = 16 , ytick={0,2,...,16} , xtick = {0,2,...,24}, xlabel = $z_1(x)$, ylabel = $z_2(x)$]

        % LUB
        \addplot[only marks,mark=square*] coordinates {(2,16) (6,14) (8,10) (10,8) (16,2) (24,1)};
        
        % Pts
        \addplot[only marks,mark = +] coordinates {(2,14) (6,10) (8,8) (10,2) (16,1)};
        % OB
        \addplot[only marks,mark=o,mark size=4pt] coordinates{(16,14)};
        
        % x_i = 0
        %\addplot[thick, dotted] coordinates{(3,14)(16,1)};
        
        % x_i = 1
        \addplot[thick, dashdotted] coordinates{(8,14)(16,6)};

        % LB
        %\addplot[thick, samples=100, domain=6:12, green,name path = A] {9 - 0.5*x};
        %\addplot[thick, samples=100, domain=4.5:6,green,name path = C] {18 - 2*x};

        % UB
        \addplot[dashed, samples=100] coordinates{(2,16)(2,14)(6,14)(6,10)(8,10)(8,8)(10,8)(10,2)(16,2)(16,1)(24,1)};

        % OB fill
        \path[fill = gray,opacity = 0.2] (0,14) -- (16,14) -- (16,0) -- (0,0);
        
        % UB fill
        %\path[fill = red,opacity = 0.1] (0,100) -- (10,100) -- (10,90) -- (30,90) -- (30,60) -- (50,60) -- (50,50) -- (70,50) -- (70,30) -- (80,30) -- (80,24) -- (90,24) -- (90,10) -- (130,10) -- (130,0) -- (0,0);

        \addlegendentry{$\NU$}
        \addlegendentry{$\UB$}
        \addlegendentry{OB}
        %\addlegendentry{$x_i = 0$}
        \addlegendentry{$x_i = 1$}
    %\addlegendentry{OB}

    \end{axis}
    \end{tikzpicture}
%\end{center}
}
            %\input{tikzFigures/VarFixDomi1}
            %\caption{The linear program $F(i,1,\lambda)$ was solved, and the optimal value $z^{*1}$ was obtained. All points of the objective space satisfying the objective branching constraints and $\lambda z(x) = z^{*1}$ are depicted by the dashdotted line. All feasible points are located both in the gray area (objective branching constraints) and above the dashdotted line(weighted sum value). Since no local upper bounds is located in this region, $x_i = 1$ is not a possible value for new non-dominated points in this region of the objective space.}
        \end{minipage}
        \caption{Both $F(i,0,\lambda)$ and $F(i,1,\lambda)$ are solved with $\lambda = (1,1)$, resulting in the situations depicted in the left and right figures respectively. Given the objective branching constraints, it is concluded that $x_i$ cannot take value $1$ and thus, $x_i$ is fixed to $0$.}
        \label{fig:varFixDomi}
    \end{figure}
    
    This strategy %approach 
    is closer to the one proposed by \cite{adelgren2021branch} for the bi-objective case in the sense that we also compute a lower bound set, namely the linear relaxation of a weighted-sum scalarization. However, our lower bound set is weaker than theirs (linear relaxation), but requires only one linear program to be solved.
    
    %It is interesting to note that % can we loose domi by avoid solving LP using preprocess rules for feasibility ???
    
    % ---------------------------------------
    %   Cover cuts
    % ---------------------------------------

    \subsection{Objective branching based cover inequalities}
    \label{sec:cc}
    
    For an objective $k$ that is minimized, an objective branching constraint has the form $w^Tx \leq b$, where $w$ is given by the objective coefficients of $z_k(x)$, and $b$ is derived from the bound on objective $k$ in the sub-problem at hand. One can observe that this constraint is in fact a knapsack constraint from which cover inequalities can be derived (see, e.g., \citet{gu1998lifted}).
    
    %, for which well-known valid inequalities exist. \comment[id=NF]{ref for this statement?}
    
    %Let $z(x) \leqq s$, $s \in \mathbb{Z}^p$, be an objective branching constraint.
    %By nature, 
    Objective branching constraints are often expected to be binding constraints, as they are included to create disjoint sub-problems. Hence, for an objective $k$ such that the objective branching constraint $z^{k} \leq s^k$ is binding, there will be extreme points in the lower bound set whose $k^{th}$ component will be equal to $s^k$. In case such an extreme point is fractional, cover cuts can be generated to cut it from the lower bound set with the aim to move it closer to an integer point.%, hoping to reach an integer point faster.
    
    Let $l \in \LB(\eta)$ be an extreme point such that $l_k = s_k$, and $x^l \in \X^{LP}(\eta)$ its pre-image. We define $\mathcal{J}^{max}(l) = \{j \in \{0,...,n\} : x^l_j = 1 \}$ as the set of indices of the variables that take the maximum value for $x^l$, i.e. value $1$. Similarly, we define $\mathcal{J}^{mid}(l) = \{j \in \{0,...,n\} : 0 < x^l_j < 1\}$ as the set of indices of the variables that take a fractional value in $x^l$, i.e. in the middle of the possible integer values. By definition, $\sum_{j \in \mathcal{J}^{max}} c^k_jx_j + \sum_{j \in \mathcal{J}^{mid}} c^k_jx_j = s^k$ holds true. If $x^l$ is fractional, i.e., $\mathcal{J}^{mid} \ne \emptyset$, then $\sum_{j \in \mathcal{J}^{max}} c^k_j + \sum_{j \in \mathcal{J}^{mid}} c^k_j > s^k$ also holds true because for all $j \in \mathcal{J}^{mid}$, we have $x_j < 1$. Hence, all variables $x_j$ such that $j \in \mathcal{J}^{max} \cup \mathcal{J}^{mid}$ cannot simultaneously take value $1$. Thus, $\sum_{j \in \mathcal{J}^{max} \cup \mathcal{J}^{mid}} x_j \leq |\mathcal{J}^{max} \cup \mathcal{J}^{mid}| - 1$ is an example of a cover inequality that can be generated from $x^l$.
    
    Of course, in many cases, different cover inequalities can be generated. These cuts can also be strengthened %, e.g., 
    by using any of the available lifting procedures from the literature. %, and all the progress from the Knapsack literature can be used for this purpose.

\section{Node selection rules}
\label{sec:nodeSel}

In the MOBB literature, breadth-first and depth-first are the commonly employed %classical 
node selection rules. Indeed, the fact that these rules are independent from the nature of the problem being solved constitutes a good reason to use such rules when expanding branch-and-bound methods to the multi-objective case. %, as they do not require additional work to make the algorithm perform correctly. 
However, previous studies have shown that depth-first is significantly better for some problem classes, whereas breadth-first is better for others (see e.g., \cite{Vincent2013corrections,parragh2019branch,Forget2022}). This inconsistency is problematic when building a generic solver as we do here, since it could easily lead to very poor performance in some cases. 

%Therefore, we aim to explore alternative node selection rules in the hope of finding rules that are more robust across problem classes, and that perform better than the classical depth and breadth-first strategies.

In the single objective literature, the so-called best-bound strategy (and variations thereof) has shown to be of value \citep{linderoth1999computational}.
%One way to address the issue is to find ways to get inspiration from the best-bound strategy used for in single-objective B\&B frameworks. 
Its basic principle consists in exploring first the node that has the lowest lower bound value, as it constitutes the most promising area of the decision space. Unfortunately, in the multi-objective case, it is often not a trivial task to determine which node has the best bound, since one may have a lower bound set that is better than the others in a particular region of the objective space, but worse in other regions.

In the following, %remainder of this Section, 
we propose two rules based on the best-bound principle. In Section~\ref{sec:wsRule}, we present a rule that searches for the best bound in a specific part of the objective space by using weighted-sum values. In Section~\ref{sec:gapRule}, we define a rule that is based on gap measures between lower and upper bound sets.

    % ---------------------------------------
    %   WS rule
    % ---------------------------------------
    \subsection{Weighted-sum rule}
    \label{sec:wsRule}
    
    A straightforward way %An easy way 
    to mimic the best-bound approach in a MOBB is to consider a weighted-sum scalarization, and to use the value of its linear relaxation as a measure of the quality of the lower bound set. % in each node. 
    Let $\eta$ be a node of the tree, let $\lambda$ be the weight vector used for the scalarization $P_{\lambda}(\lambda)$, and $z^*$ the optimal value of its linear relaxation $P_{\lambda}^{LP}(\lambda)$. The score $s(\eta)$ of the node $\eta$ is then given by $s(\eta) = z^*$, and the node with the lowest score is selected.
    
    Note that this rule is equivalent to a best-bound strategy using a branch-and-bound to solve the problem $P_{\lambda}$. Hence, translated into the context of MOBB, one can say that this strategy selects %develops 
    the node that is the most promising in direction $\lambda$ first.
    
    From a computational point of view, the score of a new node $\eta$ has to be calculated at its creation, which, in the present framework, requires solving a single-objective linear program, namely $P_{\lambda}^{LP}(\eta)$. However, using a simple re-ordering of the steps of Algorithm~\ref{alg:BB}, it is possible to obtain the score of $\eta$ without solving a linear program. Indeed, it is well known that all points of the non-dominated set of a multi-objective continuous linear problem correspond to an optimal solution of a weighted-sum scalarization \citep{Ehrgott05}. In our context, this implies that at node $\eta$, the score of $\eta$ can be obtained by searching for the point $l^* \in \LB(\eta)$ such that there is no other $l$ for which $\lambda l < \lambda l^*$. In other words, we search for the point of the lower bound set with the minimum weighted-sum value given the weight vector $\lambda$. Fortunately, this point is given by an extreme point of $\LB(\eta)$ (see Property~\ref{ppt:ws}), and only extreme points have to be checked. Hence, by computing the lower bound set at the creation of the node instead of when the node is selected, the score can be obtained at a very low cost. Note that whether the lower bound set is computed at the creation or at the selection of the node does not make a difference, as $P^{LP}(\eta)$ does not change.  %not change anything, 
    %as it is based on the branching constraints that will always remain the whole existence of the node.
    Furthermore, this %result 
    also holds for the update of the upper bound set, that only depends on the solutions found in the lower bound set. However, this is not true for fathoming, and in particular, fathoming by dominance. Indeed, new feasible points may be found between the creation and the selection of a node, which may allow the node to be fathomed by dominance. Hence, Lines~\ref{l:lb} and \ref{l:ub} are moved to after the creation of the node, and the computation of the score is performed as well, which results in the new framework given by Algorithm~\ref{alg:BB2}.
    
    \begin{algorithm}[tb]
	\caption{An alternative branch-and-bound algorithm for MOILPs using a best-bound strategy}
	\label{alg:BB2}
	\begin{algorithmic}[1]
		\STATE Create the root node $\eta^0$; set $\nodeList \leftarrow \{\eta^0\}$ and $\UB \leftarrow \emptyset$ 
		\WHILE {$\nodeList\ne \emptyset$}
			\STATE Select the node $\eta$ with the best score from $\nodeList$; Set $\nodeList \leftarrow \nodeList \backslash \{\eta\}$ 
			\IF {$\eta$ cannot be fathomed}
			    \STATE Split $P(\eta)$ into disjoint subproblems $P(\eta^1),...,P(\eta^h)$, and store each in a unique child node of $\eta$. 
			    \FOR{$\hat{\eta} \in \{\eta^1,...,\eta^h\}$}
			        \STATE Compute a local lower bound set for $P(\hat{\eta})$
    		        \STATE Update the upper bound set $\UB$
    		        \STATE Compute the score $s(\hat{\eta})$ for $\hat{\eta}$
    		    \ENDFOR
			\ENDIF
		\ENDWHILE
		\RETURN {$\UB$}
	\end{algorithmic}
\end{algorithm}

    To conclude, this rule is very cheap and easy to compute. However, the drawback is that it is very representative in one direction only and neglects other regions of the objective space. For example, by using the weight vector $\lambda = (1,...,1)$, the rule is likely to find good solutions that are well balanced across all objectives more rapidly than good solutions that are very good in one of the objectives but bad for other objectives. However, the actual impact on the performance is unclear, and is further studied in Section~\ref{sec:expe}.
    
    % ---------------------------------------
    %   Gap rule
    % ---------------------------------------
    \subsection{Gap measure rule}
    \label{sec:gapRule}
    
    Another way to adapt the idea of best-bound strategies to the multi-objective case is to compute a measure of the gap between the upper and lower bound set in each node. In this case, the node with the largest gap is explored first, as it describes either a promising area, or a region where very few feasible points have been discovered, and possibly many more remain to be discovered.
    
    An intuitive way to compute the gap in a given node is to compute the hypervolume of the search area. However, it is well known that it is a costly and difficult operation, particularly when three or more dimensions are considered. Hence, alternative measures are necessary. When \cite{Ehrgott07} introduced the concept of lower and upper bound sets, they also proposed a number of measures to compare the quality of lower and upper bound set. One of these measures is similar to the Hausdorff distance, and consists in computing the minimal distance between the two points from each set respectively that are the furthest away. Recently, this measure has been used by \cite{adelgren2021branch} to compute gaps between bound sets in the context of bi-objective mixed-integer branch-and-bound.
    
    In our context, at node $\eta$, the Hausdorff distance between the upper bound set and the lower bound set is given by $\max_{u \in \NU} \min_{l \in \LB(\eta)} d(u,l)$, where $d(u,l)$ is the distance between $u$ and $l$. From an implementation point of view, only local upper bounds that are located above the lower bound set are considered, as they are the only ones that define the search region in $\eta$. If there is none, we consider that the node has a gap of $0$. This approach is, in fact, analogous to the single objective case: as long as the lower and upper bound sets have not met entirely, the gap is strictly positive, and the node cannot be fathomed by dominance.
    
    When multiple objectives are considered, %a difficulty arises due to  the fact that 
    %when 
    and a new feasible point $u$ is added to the upper bound set $\UB$, the gap in some nodes may change. In particular, a node $\eta^1$ with a smaller gap than $\eta^2$ may end up with a gap larger than that of $\eta^2$. This implies that in order to identify %extract 
    the node with the best score, the gaps have to be recomputed whenever $\UB$ changes. %in all nodes when new points are found. 
    Unfortunately, this may be computationally expensive, as it is not rare that many nodes are open at a given iteration of the MOBB.
    
    To reduce the computational burden, we rely on two simple properties: %Fortunately, it is possible to reduce the computational cost of the update by combining two simple properties. 
    First, whenever a new point is added to the upper bound set, the gaps at all nodes can only stay the same or decrease. The reason is that the lower bound sets remain unchanged and  % This comes from the fact that the lower bound set always stays the same at any iteration, and 
    the upper bound set improves when a new feasible point is found. This implies that the search region shrinks: the upper bound set moves closer to the lower bound sets. %, and the lower and upper bound sets become closer to each other. 
    Second, we are only interested in the node with the largest gap, as it corresponds to the next node being explored. Let $\eta^1 \in \mathcal{T}$ be the node with the largest gap, and $\eta^2 \in \mathcal{T}$ be the node with the second largest gap. Let $g^{old}(\eta)$ be the gap of a node $\eta$ before re-computation, and $g^{new}(\eta)$ be its gap after re-computation. By construction, we know that $g^{old}(\eta^2) \geq g^{new}(\eta^2)$. Furthermore, for all $\eta \in \mathcal{T} \backslash \{\eta^1, \eta^2\}$, we have $g^{old}(\eta^2) \geq g^{old}(\eta) \geq g^{new}(\eta)$. Hence, if $g^{new}(\eta^1) \geq g^{old}(\eta^2)$, only by re-computing the gap in $\eta^1$, we know that $\eta^1$ is the node with the largest gap after update of the upper bound set. If the condition is not satisfied, $\eta^1$ is put back into $\mathcal{T}$ and the process is repeated with $\eta^2$, the new potential node with the largest gap. The selection procedure is given in Algorithm~\ref{alg:nodeSelGap}.
    
    \begin{algorithm}[tb]
	\caption{Selection of the node with the largest gap}
	\label{alg:nodeSelGap}
	\begin{algorithmic}[1]
		\STATE found $\leftarrow$ FALSE
		\WHILE {!found}
			\STATE Select the node $\eta^1$ with the largest gap from $\nodeList$; Set $\nodeList \leftarrow \nodeList \backslash \{\eta\}$
			\STATE Compute $g^{new}(\eta^1)$
			\STATE Select the node $\eta^2$ with the largest gap from $\nodeList$
			\IF {$g^{new}(\eta^1) \geq g^{old}(\eta^2)$}
			    \STATE found $\leftarrow$ TRUE
			\ELSE
			    \STATE $g^{old}(\eta^1) \leftarrow g^{new}(\eta^1)$
			    \STATE $\nodeList \leftarrow \nodeList \cup \{\eta^1\}$
			\ENDIF
		\ENDWHILE
		\RETURN {$\eta^1$}
	\end{algorithmic}
\end{algorithm}
%\comment[id=NF]{Proof of termination?}

\section{Experiments}
\label{sec:expe}

%In this section, we report experimental results with %on 
%the branch-and-bound framework. 
All algorithms are implemented in C++17, relying on %using 
Cplex 20.1 for solving single-objective linear programs, using a single thread.  %and all computations are serial. 
The experiments are carried out on Linux 10.3, on a Quad-core X5570 Xeon CPUs @2.93GHz processor and with 48GB of RAM. % memory. 
A time limit of one hour is set when running the algorithms.

%In this section, we conduct experiments with different configurations for the branch-and-bound to answer some research questions.
Our computational study aims at answering the following
%These 
research questions: % are the followings: 
(i) How does probing perform? In particular, how does it perform in combination with objective branching, and why? (ii) What is the impact of using an %the 
objective function in the linear programs used for performing probing? (iii) What is the impact of deriving cover cuts  %inequalities generated with 
from the objective branching constraints on the performance of the algorithm? (iv) Can node selection rules based on the best-bound idea outperform the classical depth and breadth-first strategies often used in the literature?  %Is there room for further improvement by investigating variable selection rules as well? 
(v) Computing lower bound sets in the multi-objective case is expensive.  Does resorting to pure enumeration at certain nodes in the tree improve the performance of the proposed MOBB?  %Would MOBB benefit from pure enumeration without lower bound computation? 
(vi) By fixing variables, the set of potential candidates for branching is reduced. What is the impact of the chosen variable selection rule? 
(vii) How does the proposed %this 
branch-and-bound framework perform in comparison to state-of-the-art % compared to recent 
objective space search algorithms?

%\begin{itemize}
%    \item How does variable fixing perform? In particular, how does it perform in combination with objective branching, and why?
%    \item What is the impact of the objective function in the linear programs used for variables fixing?
%    \item What is the impact of cover cuts inequalities generated with objective branching constraints on the performance of the algorithm?
%    \item Can node selection rules based on the best-bound idea outperform the classical depth and breadth first strategies often used in the literature?
%    \item By fixing variables, the set of potential candidates for branching is reduced. Is there room for further improvement by investigating variable selection rules as well? % By fixing variables, the set of potential candidates for branching is reduced. Given this fact, i
%    \item How does this branch-and-bound framework performs compared to recent objective space search algorithms?
%\end{itemize}

%The tests are carried over a set of:
We test our algorithms on the following three different types of problems and benchmark instances:
\begin{itemize}
    \item Capacitated Facility Location Problem (CFLP). The instances are taken %extracted 
    from \cite{DuleabomCFLP}. Instances with 3 objectives 65, 230, and 495 variables are considered.
    \item Knapsack Problem (KP). The instances from \cite{Moolibrary-Kirlik14} are used. Instances with 3 objectives and 40, 50, 60, 70, 80 variables are solved, as well as instances with 4 objectives and 20, 30, 40 variables.
    \item Uncapacitated Facility Location Problems (UFLP). The instances are extracted from \cite{Forget2022}. For 3 objectives, instances with 56, 72, 90, 110 variables are used. For 4 objectives, instances with 42 and 56 variables are solved.
\end{itemize}

For each problem class, number of objectives, and number of variables, 10 instances are solved, leading to a total of 170 instances. % tested.

%Various configurations are tested for the branch-and-bound algorithm in order to address the different research questions raised (objective branching, variable fixing, generation of cover cuts inequalities, node selection). The configurations tested will be clearly stated together with the results of each experiment. 
Unless specified otherwise, the framework uses the following parameters and heuristics:

\begin{itemize}
    \item \textbf{Lower bound sets:} the linear relaxation is used as lower bound set. Its computation is warm-started by using the algorithm from \cite{Forget2022};
    \item \textbf{Local upper bounds:} the set of local upper bounds $\NU$ is updated whenever a new point is added to the upper bound set $\UB$ by using the algorithm from \cite{Klamroth15}. Furthermore, all objective functions are expected to have integer coefficients, and all variables are binary. This implies that the non-dominated points can take integer values only. Consequently, when performing the dominance test and computing sub-problems through objective branching, each component of the local upper bound is shifted by $-1$.
    \item \textbf{Variable selection rule:} At the creation of sub-problems in the decision space, a free variable is chosen for branching. First, this %a different 
    variable is chosen independently in each sub-problem obtained from %created using 
    objective branching. Let $s \in \Rp$ be the super local upper bound defining the sub-problem in which a free variable has to be chosen. The variable that is the most often fractional among the extreme points $l$ of the lower bound set $\LB(\eta)$ that satisfies $l \leqq s$ is selected. In case of ties, the one whose average value is closest to $0.5$ is selected and in case of equal average values,  %If multiple variables can be considered as the most often fractional,  the one with the average value the closest to $0.5$ among them is selected. 
    %Finally, if 
    %there are still multiple candidates for branching, 
    the one with the smallest index is chosen.
\end{itemize}

Furthermore, unless specified otherwise, cover cut generation is disabled. %Then, a 
A number of different configurations are tested. The three main components evaluated in this study are the following:

\begin{itemize}
    \item \textbf{Objective branching:} three options are considered: no objective branching (\texttt{NOB}); cone bounding (\texttt{CB}); and full objective branching (\texttt{FOB}), as presented in Section~\ref{sec:BB}. Cone bounding is an alternative to objective branching proposed by \cite{Forget20a}. The idea is to derive upper bounds on the objective functions from the partial dominance of the lower bound set, but without splitting the objective space into sub-problems, i.e., only decision space branching is performed. 
    %First, no objective branching is used at all (\texttt{NOB}). Then, cone bounding (\texttt{CB}), an alternative to objective branching proposed by \cite{Forget20a}. The idea is to derive upper bounds on the objective functions from the partial dominance of the lower bound set, but without splitting the objective space into sub-problems, i.e., only decision space branching is performed. Finally, the third configuration is full objective branching (\texttt{FOB}), as presented in Section~\ref{sec:BB}.
    \item \textbf{Probing/variable fixing:} %{Variable fixing:} 
    three possibilities are considered: no variable fixing (\texttt{NVF}); variable fixing using the advanced strategy % with no objective function 
    as presented in Section~\ref{sec:vfAdvStrat} (\texttt{VF}); and variable fixing using a weighted-sum objective function to allow for variable fixing by dominance %the dominance rule
    (\texttt{VFD}) as explained in Section~\ref{sec:objFunc}.
    \item \textbf{Node selection rule:} four configurations are considered:
    best of depth-first and breadth-first (\texttt{DB}); best-bound based on weighted sums (\texttt{BBWS}); a normalized version (\texttt{BBWSN}); and best-bound based on the gap measure presented in Section~\ref{sec:gapRule} (\texttt{BBGAP}). In the case of (\texttt{DB}), the best strategy per problem class is used. CFLP and UFLP use breadth-first, whereas KP uses depth-first. \texttt{BBWS} and \texttt{BBWSN} use $\lambda = (1,...,1)$ (see Section~\ref{sec:wsRule}). Normalization may be important for problems for which the coefficients of the different objective functions take values in very different ranges, such as the CFLP.
    %The first one is depth-first or breadth-first  (\texttt{DB}). For each problem class, only the best of the two with respect to CPU time is used. CFLP and UFLP use breadth-first, whereas KP uses depth-first. The second configuration uses the best-bound idea based on the weighted-sum value from Section~\ref{sec:wsRule} (\texttt{BBWS}), using a weight of $(1,...,1)$. The third configuration is a variant of \texttt{BBWS} where objective functions are normalized by the range of the coefficients of their objective function (\texttt{BBWSN}). The idea behind this is to have a weight vector that is more balanced towards all the objectives. This will be important in particular for CFLP, for which the coefficients of the different objective functions take values in very different ranges. Finally, the fourth configuration is the best-bound idea based on the gap measure presented in Section~\ref{sec:gapRule} (\texttt{BBGAP}).
\end{itemize}

In the remainder of this section, each sub-section is designed to address one of the research questions raised earlier.

    % ---------------------------------------
    %   Question 1
    % ---------------------------------------
    
    \subsection{Probing and objective branching}
    
    In a first step, %Here, 
    we investigate the effect of combining objective branching and probing. We fix the node selection rule to configuration \texttt{DB}.
    % We fix the node selection rule to breadth-first for CFLP and UFLP, and depth-first for KP (best performance between depth and breadth first for each problem class). All cut generation are disabled.
    %First, 
    Six configurations are tested: the %all 
    three objective branching strategies (\texttt{NOB}, \texttt{CB}, and \texttt{FOB}), in combination with (\texttt{VF}) and without (\texttt{NVF}) variable fixing. %combined 
    %with two variable fixing configurations (\texttt{NVF} and \texttt{VF}). 
    For \texttt{FOB} and \texttt{CB}, probing is performed only when an improvement is observed in the objective branching constraints, as suggested in Section~\ref{sec:vfAdvStrat}. However, probing is performed at %for
    every node for \texttt{NOB}.
        
    \begin{figure}
        \centering
        \resizebox{\linewidth}{!}{\input{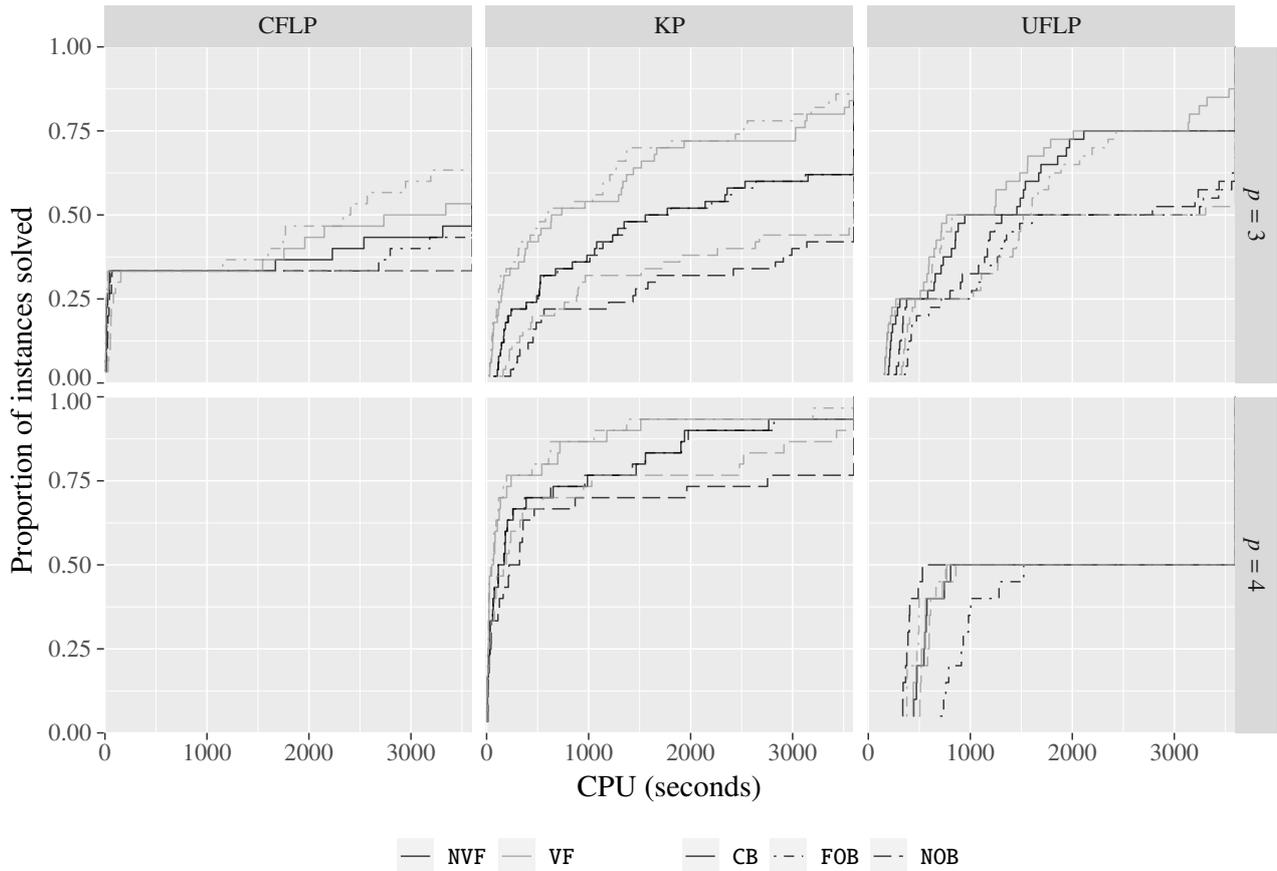}}
        \caption{Performance curves of the different objective branching and probing %variable fixing 
        configurations. The x-axis represents the CPU time expressed in seconds, and the y-axis corresponds to the proportion of instances solved. %Each curve represents the proportion of instances solved through time for a given configuration. 
        The line type indicates %corresponds to 
        the objective branching configuration %used, 
        and the line color %to 
        the variable fixing configuration.}
        \label{fig:PerformanceProfileVarfix}
    \end{figure}
    
    In Figure~\ref{fig:PerformanceProfileVarfix}, the performance profiles of the different configurations are %is 
    given. The x-axis represents the CPU time expressed in seconds, whereas the y-axis corresponds to the proportion of instances solved. From this figure, it is clear that for both CFLP and KP, the winning configuration is \texttt{FOB-VF}. For UFLP and $p = 3$, however, although \texttt{FOB-VF} is the best configuration for smaller instances, \texttt{CB-VF} becomes the winning configuration for larger instances. For UFLP and $p = 4$, \texttt{NOB-NVF} is slightly faster than the other configurations.
    % A similar observation was made in the experiments of \cite{Forget20a} \comment[id=NF]{... when the paper come out!}
    
    It is interesting to note that enabling probing resulted in the greatest speed-ups for full objective branching (\texttt{FOB}). This is particularly striking for UFLP, where \texttt{FOB-NVF} is among the worst configurations, but \texttt{FOB-VF} is competitive with the best configurations. For CFLP, it is clear that \texttt{FOB} benefits more from probing than \texttt{CB} or \texttt{NOB}, since \texttt{CB-NVF} is faster than \texttt{FOB-NVF}, but \texttt{FOB-VF} is faster than \texttt{CB-VF}. This suggests that the branch-and-bound algorithm benefits the most from probing when tight objective branching constraints (i.e., small sub-problems) are generated.
    
    Finally, it appears from Figure~\ref{fig:PerformanceProfileVarfix} that when no objective branching is used (\texttt{NOB}), probing (\texttt{VF}) can be slightly beneficial (KP) but also worsen the performance of the algorithm (CFLP, UFLP). This implies that probing is efficient mainly \emph{in combination} with objective branching in MOBB.
    
    %\begin{table}[]
    %    \centering

    %    \input{tablesExpe/tabVarFixNodes}
    %    \caption{The average number of nodes explored over 10 instances for each problem class, number of objectives, number of variables, and configuration. The number in brackets is the number of instances unsolved. Note that when the number of unsolved instances is high, the number of nodes explored may be low due to the fact that the algorithm could not explore a large number of nodes within the time limit of one hour.}
    %    \label{tab:varFixNodes}
    %\end{table}
    
    The speed-ups can be largely explained by looking at the size of the branch-and-bound tree.
    %In Table~\ref{tab:varFixNodes}, the average number of nodes explored is given. The first observation is that
    Indeed, the number of nodes explored is positively correlated to the CPU time (see Table~\ref{tab:varFixNodes} in the appendix for detailed performances). For \texttt{NOB}, small differences are observed between \texttt{NVF} and \texttt{VF} in terms of the number of nodes, whereas larger gaps are observed for \texttt{CB}. %Finally, 
    Similarly to the CPU time, the most significant differences between \texttt{NVF} and \texttt{VF} are observed for \texttt{FOB}. 
    Indeed, if we consider all instances for which both \texttt{FOB-NVF} and \texttt{FOB-VF} are solved, \texttt{FOB-VF} resulted in %developed 
    $12.78$ times fewer %less 
    nodes than \texttt{FOB-NVF} for $p = 3$.
    In many cases, \texttt{FOB-NVF} is the configuration with the largest tree size, whereas \texttt{FOB-VF} has the smallest. This indicates that probing strongly helps to reduce the size of the tree, both by improving lower bound sets and helping the algorithm to make better branching decisions. Note that a similar observation was made by \cite{adelgren2021branch} for the bi-objective case.
    
    Finally, one may notice that the speed-ups in terms of CPU times are smaller than the gains in terms of the number of nodes explored. This is due to the fact that probing has a significant cost: on average, $23.5\%$ and $40.36\%$ of the total CPU time for \texttt{CB} and \texttt{FOB} respectively, over all instances (see Table~\ref{tab:apdxProp} in the appendix for more details).
    
    % ---------------------------------------
    %   Question 1 bis
    % ---------------------------------------
    
    \subsection{Combining probing and bounding}
    %Impact of the objective function when performing probing}
    
    We %are 
    now analyze %interested in 
    the impact of introducing a weighted-sum %of the objectives as an 
    objective function when performing probing (setting \texttt{VFD}), as suggested in Section~\ref{sec:objFunc}. It is important to test both cases (\texttt{VF} and \texttt{VFD}) because, on the one hand, introducing an objective function improves the fathoming potential, but, on the other hand, may also result in linear programs that are %becoming 
    more difficult to solve. For this purpose, we now consider only two configurations: \texttt{FOB-VF} and \texttt{FOB-VFD}, using as weight vector % of 
    $\lambda = (1,...,1)$ %is used for 
    in the weighted-sum %of the 
    objective. % in \texttt{FOB-VFD}.
    %Two configurations are tested, namely varFix and varFixDomi. In varFix, the objective function $\min 0$ is used, whereas $\min \lambda z(x)$ is used in varFixDomi. The weight-vector used is $(1,...,1)$ after normalizing the objective function by the range of their coefficient. The idea is to have a weight vector that is balanced over all objective functions.
    
    \begin{table}[]
        \centering
        \resizebox{\linewidth}{!}{
        
\begin{tabu} to 1.4\linewidth {>{\centering}X>{\centering}X>{\centering}X>{\centering}X>{\centering}X}
\toprule
p & pb & n & \texttt{FOB-VF} & \texttt{FOB-VFD}\\
\midrule
 &  & 65 & 10.7 & 10.4\\

 &  & 230 & 2332.7 & 2277.7\\

 & \multirow{-3}{*}{\centering\arraybackslash CFLP} & 495 & 3601.3 & 3601.2\\
 
 \cmidrule{2-5}

 &  & 40 & 65.8 & 64.6\\

 &  & 50 & 172.4 & 167.7\\

 &  & 60 & 749.7 & 693.9\\

 &  & 70 & 2201.7 & 2033.3\\

 & \multirow{-5}{*}{\centering\arraybackslash KP} & 80 & 3140.0 & 3048.7\\
 
 \cmidrule{2-5}

 &  & 56 & 197.0 & 183.5\\

 &  & 72 & 677.6 & 619.7\\

 &  & 90 & 1934.4 & 1805.7\\

\multirow{-12}{*}{\centering\arraybackslash 3} & \multirow{-4}{*}{\centering\arraybackslash UFLP} & 110 & 3600.1 & 3600.1\\
\cmidrule{1-5}
 &  & 20 & 8.8 & 8.9\\

 &  & 30 & 53.9 & 53.7\\

 & \multirow{-3}{*}{\centering\arraybackslash KP} & 40 & 1138.0 & 1138.2\\
 
 \cmidrule{2-5}

 &  & 42 & 489.2 & 486.9\\

\multirow{-5}{*}{\centering\arraybackslash 4} & \multirow{-2}{*}{\centering\arraybackslash UFLP} & 56 & 3600.2 & 3600.1\\
\bottomrule
\end{tabu}
        }
        \caption{Average CPU time expressed in seconds over 10 instances for each problem class, number of objectives $p$, and number of variables $n$. Two configurations are tested: \texttt{VF} and \texttt{VFD}, both in combination with \texttt{FOB}.}
        \label{tab:varFixDomiCPU}
    \end{table}
    
    %The performance of the two configurations is given in Table~\ref{tab:varFixDomiCPU}. 
    Table~\ref{tab:varFixDomiCPU} gives the average CPU time of the two configurations per instance class.
    On average, it appears that the objective function has a minimal 
    impact on the CPU time. Except in rare cases (e.g., KP for $p = 4$), introducing the objective function is still slightly beneficial. A speed-up of $4.64\%$ is observed on average across all instances for which both \texttt{FOB-VF} and \texttt{FOB-VFD} are solved.
    
     A potential explanation for the low gain is %comes from the fact 
     that the linear relaxation of a weighted-sum scalarization provides a rather weak lower bound set. This is true in particular in the case where $p \geq 3$, as many local upper bounds can have %take 
     infinite components. Such local upper bounds also have an infinite weighted-sum value and therefore, do not allow for variable fixing by dominance. %are thus always dominated by the lower bound set obtained here. 
     In the bi-objective case, in comparison, at most, two local upper bounds can take infinite components. % in the bi-objective case. 
     One could imagine a strategy where, if a local upper bound has %with 
     an infinite value at component $k$ %is discovered 
     in the considered node-problem, the weight of objective $k$ is set to $0$. %sub-problem at hand, 
     %a value of $0$ is set in component $k$ of the weight vector $\lambda$. 
     %However, the lower bound set then becomes rather weak, as it is more difficult to reach the local upper bounds that take a large value on the remaining objectives.
     This strategy was tested in preliminary experiments, but similarly to the weight vector $(1,...,1)$, it led to small speed-ups only. 
    
    % ---------------------------------------
    %   Question 3
    % ---------------------------------------
    
    \subsection{Node selection rules}
    
    We now investigate the performance of the node selection rules proposed in Section~\ref{sec:nodeSel}. In this section, we use \texttt{FOB-VF} as the base configuration, and combine it with four node selection strategies%are tested
    : \texttt{DB}, \texttt{BBWS}, \texttt{BBWSN}, and \texttt{BBGAP}.
    
    \begin{figure}
        \centering
        \resizebox{\linewidth}{!}{\input{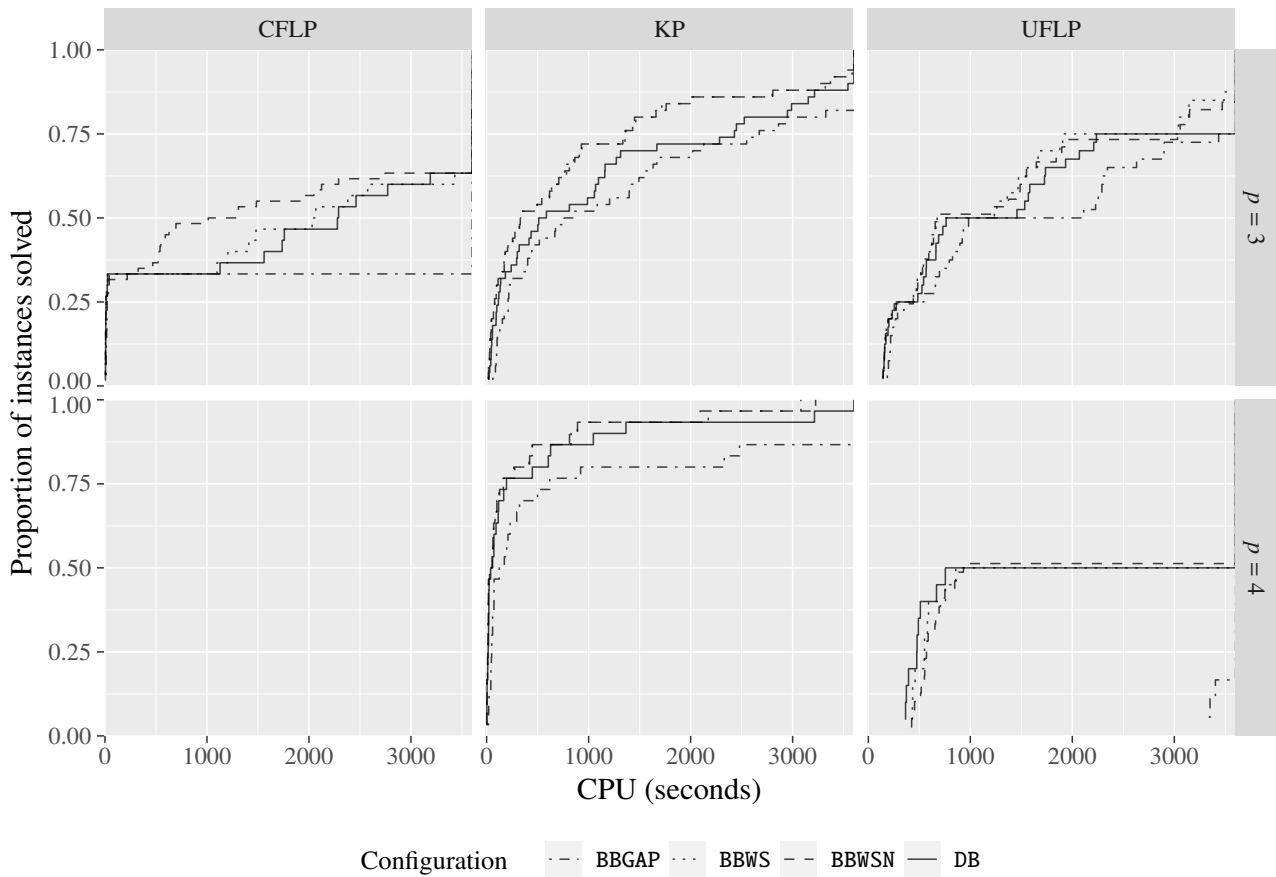}}
        \caption{Performance curves of the different node selection configurations in combination with \texttt{FOB-VF}. The x-axis represents the CPU time expressed in seconds, and the y-axis corresponds to the proportion of instances solved. %Each curve represents the proportion of instances solved through time for a given configuration.
        }
        \label{fig:nodeSelCPU}
    \end{figure}
    
    The results are given in Figure~\ref{fig:nodeSelCPU}. The x-axis represents the time elapsed in seconds, and the y-axis corresponds to the proportion of the instances solved. Hence, each curve represents the proportion of instances solved over time for each configuration. The first observation is that in most cases, \texttt{BBWS} performs better than \texttt{DB}, and similarly in the worst cases. %This is, in fact, a promising result because \texttt{BBWS} both does not} %did not 
   % require any fine-tuning of the node selection rule depending on the problem class, and performed better on average. Indeed, 
   As explained earlier, configuration \texttt{DB} uses depth-first for KP and breadth-first for CFLP and UFLP, and it is not rare to see either depth-first or breadth-first used in the literature. However, a problem dependent node selection rule %fine-tuning the node selection rule depending on the problem class 
   is not desirable in the context of a generic solver such as MOBB.
   %a desirable property when building a generic solver such as a MOBB, since the solver is not supposed to know the class of the problem being solved. Moreover, preliminary experiments showed that for some problem classes, there was a huge gap in performance between depth-first and breadth-first strategies. 
   %Hence, 
   %The fact that \texttt{BBWS} performs better on average} than \texttt{DB} overcomes this significant drawback, and it even} increases the efficiency of the MOBB framework. % as well.
   Using \texttt{BBWS} instead of depth- or breadth-first overcomes this significant drawback, and it even increases the efficiency of the MOBB framework: \texttt{BBWS} performs better on average than \texttt{DB}. % as well.
    
    For KP and UFLP, the base version (\texttt{BBWS}) and the normalized version (\texttt{BBWSN}) perform similarly. This is expected, since all objective coefficients take values in the same range for each objective function. % in both problem classes. 
    This is, however, not the case for CFLP, where an interesting difference in the performance of \texttt{BBWS} and \texttt{BBWSN} is observed: % In particular, 
    \texttt{BBWSN} is faster than \texttt{BBWS}, which suggests that normalization is important whenever heterogeneous objective functions are considered, as it is usually the case when dealing with real-world problems. %it is more beneficial to consider a weight vector that does not favor some of the objectives.
    
    Finally, \texttt{BBGAP} is the worst configuration, despite exploring on average $45.2\%$ fewer nodes than \texttt{BBWSN} over all instances solved for both configurations (see Table~\ref{tab:apdxGap} in the Appendix for more details).
    %PLEASE PROVIDE TABLE IN APPNDIX AND REFER TO THIS TABLE HERE}.
    This is due to the heavy computational cost of computing the gap measures in each node. Indeed, updating the gap measure when selecting a node (Algorithm~\ref{alg:nodeSelGap}) uses, on average, $17.37\%$ of the total CPU time for $p = 3$, and $36.37\%$ for $p = 4$ (see Table~\ref{tab:apdxGap} in the Appendix for more details).
    %PLEASE PROVIDE TABLE FOR THIS VALUES IN APPENDIX}.
    Combined with the fact that the gap at each node has to be computed at its creation as well, the %total cost of 
    additional computation time required for computing the gap measure %this node selection 
    rule outweighs %overwhelms 
    the benefit of exploring fewer nodes, overall  %gain obtained from the reduction of the number of nodes, 
    leading to worse CPU times. Furthermore, the reduction in the number of nodes comes largely from those %the 
    nodes which have %with 
    a gap of 0 at %the 
    creation. Those nodes are discarded immediately and not counted among the number of processed nodes. In all other settings, they are put into the list of open nodes $\nodeList$ and are then fathomed by dominance once they are processed.
    %Indeed, in such nodes the lower bound set is dominated by the upper bound set and thus, they are discarded from the search. However, these nodes are usually fast to process, as the linear relaxation is already computed and no probing is performed, meaning that the gain in terms of CPU time obtained by skipping these nodes is rather small.} %This, however, suggests that lower} %better 
    %CPU times could possibly be achieved if a less computationally heavy gap measure is used.
    
    % ---------------------------------------
    %   Question 2
    % ---------------------------------------
    
    \subsection{Objective branching and cover cuts}
    
    In this section, we investigate the incorporation of cover cuts %inequalities generated 
    derived from the objective branching constraints. We focus on UFLP, as all of its objectives are %in 
    minimization objectives. They are constructed following the idea presented in Section~\ref{sec:cc}, and lifted using the procedure from \cite{Letchford_2019}.
    
    In preliminary experiments, we tested to resolve the linear relaxation and regenerate cover cuts on the extreme points until no new cut could be generated. However, in the multi-objective case, cutting an extreme point in the objective space implies that one or several facets and extreme points are generated, which %itself 
    requires as many linear programs to be solved when using a Benson-type algorithm. In the end, the stronger lower bound sets clearly did not compensate for the larger number of linear programs solved. Hence, we opted for a less expensive strategy.
    
    In this alternative cut-generation scheme, cover cuts are generated after objective branching is computed and before performing probing, with the expectation that the newly generated cuts will help fixing variables.
    Let $s$ be the super local upper bound that defines the objective branching constraints, cover cuts are generated at each node for each objective bounded by a constraint, and for %on 
    each extreme point $l$ of the lower bound set that has non-integer pre-images. 
    %and that satisfies $l \leqq s$.
    %Note that since cover cuts are generated after objective branching is performed,} %computed, 
    %not all extreme points of the lower bound set necessarily satisfy $l \leqq s$.} %WHY IS $l \leqq s$ EXPLICITLY MENTIONED? SHOULDN'T THEY ALL SATISFY THIS? A CUT SHOULD PROBABLY BE GENERATED IF $l_k = s_k$, I.E., WHEN THE CONSTRAINT IS BINDING. IS THIS WHAT IS HAPPENING HERE?}.
    %COVER CUTS SHOULD ONLY BE GENERATED IF THEY ARE VIOLATED - IS THIS WHAT THE NEXT TWO SENTENCES TRY TO SAY?} Note that it often occurs that no cover cut is generated on an extreme point, for example, if the point is located too far from any of the objective branching constraints, or if the newly generated cut is redundant with an already existing one.
    Moreover, a cut is generated only if it is violated, i.e. if it cuts the pre-image of an extreme point of the lower bound set from the feasible set. Naturally, cuts are more likely to be generated on extreme points $l$ that have one component that is close to the objective branching constraint, i.e., for a $k \in \{1,...,p\}$ where $l_k$ is close to $s_k$.
    
    Since the linear relaxation is not resolved, no additional LP is expected to be solved in the current node. However, the generated cuts are kept in the child nodes.
    %I THINK THAT SOME MORE INFORMATION ON HOW THE CUTS ARE ADDED AND IF THEY STAY ACTIVE AND SO ON IS NECESSARY HERE. USUALLY, CUTS ARE GENERATED AND THEN THE LP RELAXATION IS SOLVED AGAIN UNTIL NO MORE VIOLATED CUTS CAN BE GENERATED. THEREFORE CUT GENERATION WILL ALWAYS IMPLY MORE LPS AT EACH NODE BUT FEWER NODES. WHAT IS EXACTLY DONE HERE?}
    % Similarly to variable fixing, when enhancing the objective branching constraints defined by a super local upper bound $s$, c
    
    %The expectation is to be able to fix more variables using the cover inequalities generated, and to strengthen the lower bound set in the children nodes.

    We compare
    %In this section, we consider only 
    the best previous configuration, namely \texttt{FOB-VF-BBWSN}, %. Then, two configurations are tested: 
    with (\texttt{CC}) and without cover cuts (\texttt{NCC}).
    \begin{table}[]
        \centering
        %\begin{table}
%\centering
\resizebox{0.65\linewidth}{!}{
\begin{tabular}{rrrrrrrrrr}
\toprule
\multicolumn{2}{c}{ } & \multicolumn{2}{c}{CPU} & \multicolumn{2}{c}{\# nodes} & \multicolumn{2}{c}{\# unsolved} & \multicolumn{2}{c}{\# LP (LP relax)} \\
\cmidrule(l{3pt}r{3pt}){3-4} \cmidrule(l{3pt}r{3pt}){5-6} \cmidrule(l{3pt}r{3pt}){7-8} \cmidrule(l{3pt}r{3pt}){9-10}
p & n & \texttt{NCC} & \texttt{CC} & \texttt{NCC} & \texttt{CC} & \texttt{NCC} & \texttt{CC} & \texttt{NCC} & \texttt{CC}\\
\midrule
 & 56 & 177.3 & 168.4 & 12282.8 & 11275.4 & 0 & 0 & 252691.1 & 249869.1\\

 & 72 & 564.4 & 593.5 & 26574.4 & 23447.7 & 0 & 0 & 634866.9 & 648676.7\\

 & 90 & 1590.5 & 1712.5 & 52141.6 & 50846.0 & 0 & 0 & 1395326.9 & 1487256.3\\

\multirow{-4}{*}{\raggedleft\arraybackslash 3} & 110 & 3381.7 & 3600.0 & 80985.7 & 45316.3 & 5 & 10 & 2530157.0 & 2124939.1\\
\cmidrule{1-10}
 & 42 & 558.9 & 668.5 & 22393.8 & 21351.6 & 0 & 0 & 688906.2 & 730118.7\\

\multirow{-2}{*}{\raggedleft\arraybackslash 4} & 56 & 3600.0 & 3600.0 & 39079.1 & 32976.8 & 10 & 10 & 2536207.7 & 2612500.7\\

\bottomrule
\end{tabular}}
%\end{table}
        \caption{Comparison of the setting \texttt{FOB-VF-BBWSN} with (\texttt{CC}) and without (\texttt{NCC}) cover cut generation for % Only results for 
        UFLP. % are shown in this table. 
        For each number of objectives and number of variables, it reports the average CPU time (CPU), the average number of nodes explored (\# nodes), the number of instances unsolved (\# unsolved), and the average total number of linear programs solved to compute the linear relaxation (\# LP (LP relax)).}
        \label{tab:cc}
    \end{table}
    The results of this experiment are given in Table~\ref{tab:cc}. Configuration \texttt{CC} generally performs slightly worse in terms of CPU time and instances solved than \texttt{NCC}, even though the number of nodes is in general slightly lower for \texttt{CC}. % (see Table~\ref{tab:cc}).}
    %This result can be surprising (BUT MORE LPS IS NOT SUPRISING TO ME...!)} %a surprise 
    %when looking at the number of nodes in Table~\ref{tab:cc}, as it is generally slightly lower for \texttt{CC}. This suggests that the cuts come with a hidden cost that is not not entirely} compensated by} %enough with 
    %the decrease in the size of the tree. 
    
    The performance of \texttt{CC} can be understood by looking at the last column of Table~\ref{tab:cc}, where the average total number of linear programs solved to compute the lower bound sets in an entire tree is reported. It tends to increase when cover cuts are used, which implies that after generating cuts, the lower bound sets in the child nodes are more complex, and require more linear programs to be solved for \texttt{CC}. This observation makes sense as the newly generated constraints are kept in the child nodes.
    %Furthermore, when computing the lower bound set, these constraints are expected to often be binding as they cut an extreme point from the lower bound set in the parent} %father 
    %node.
    %I THINK THAT THIS IS NOT THE GOOD EXPLANATION - SEE MY COMMENTS ABOVE - CUT GENERATION USUALLY IMPLIES THAT MORE LPS NEED TO BE SOLVED AT EACH NODE.}
    This possibly results in more facets to generate and, consequently, more computational effort, even though the linear relaxation is not resolved after generating cuts. % required. 
    This phenomenon highlights a difficulty of cut generation in the multi-objective case: even simple cuts can become a  computational burden, because they can easily complexify the lower bound set, possibly resulting in a considerable %huge 
    additional effort to obtain only small improvements.
 
 % ---------------------------------------
    %   Question 4 bis
    % ---------------------------------------
    
    \subsection{Enumeration in MOBB}
    
    In the deepest parts of the tree, only a few variables are free. Consequently, the lower bound sets are also expected to be very simple, i.e., made of a few facets and extreme points only. According to \cite{Forget2022}, each facet and extreme point possibly requires solving one linear program. Given the fact that the number of free variables is low, one may enumerate all possible solutions and update the upper bound set accordingly instead of keeping branching and computing lower bound sets until all nodes are fathomed. Here, after performing probing, if less than $14$ variables are free, all $2^{14} = 16384$ solutions are enumerated. This value has been determined with preliminary experiments on a sub-set of instances.
    
    Results are reported in Table~\ref{tab:OSSenum}. Configuration \texttt{BB} corresponds to the best generic branch-and-bound so far, i.e., it uses full objective branching (\texttt{FOB}), variable fixing with an objective function (\texttt{VFD}), and the best-bound node selection rule based on the normalized weighted-sum idea (\texttt{BBWSN}). Configuration \texttt{BB-E} uses the same parameters, except that enumeration as described above is enabled.
    
    \begin{table}[]
        \centering
        \resizebox{\linewidth}{!}{
        
\begin{tabu} to 1.4\linewidth {>{\centering}X>{\centering}X>{\centering}X>{\centering}X>{\centering}X>{\centering}X>{\centering}X}
\toprule
\multicolumn{3}{c}{ } & \multicolumn{2}{c}{CPU} & \multicolumn{2}{c}{\# LP} \\
\cmidrule(l{3pt}r{3pt}){4-5} \cmidrule(l{3pt}r{3pt}){6-7}
p & pb & n & \texttt{BB} & \texttt{BB-E} & \texttt{BB} & \texttt{BB-E}\\
\midrule
 &  & 65 & 10.0 & 10.0 & 44239.6 & 43403.2\\

 &  & 230 & 1988.0 & 1993.3 & 4159912.0 & 4179247.3\\

 & \multirow{-3}{*}{\raggedright\arraybackslash CFLP} & 495 & 3600.0 & 3600.0 & 2795418.2 & 2920081.0\\
 
 \cmidrule{2-7}

 &  & 40 & 46.7 & 42.4 & 277151.0 & 249235.8\\

 &  & 50 & 116.5 & 108.2 & 668678.7 & 620524.5\\

 &  & 60 & 457.8 & 431.0 & 2517589.2 & 2381897.1\\

 &  & 70 & 1414.1 & 1341.1 & 7412250.4 & 7037987.6\\

 & \multirow{-5}{*}{\raggedright\arraybackslash KP} & 80 & 2491.5 & 2388.9 & 12508369.2 & 12154795.8\\
 
 \cmidrule{2-7}

 &  & 56 & 177.3 & 134.1 & 617053.7 & 532883.8\\

 &  & 72 & 564.4 & 421.3 & 1653259.3 & 1522324.6\\

 &  & 90 & 1590.5 & 1229.2 & 4074781.8 & 3860124.8\\

\multirow{-12}{*}{\raggedleft\arraybackslash 3} & \multirow{-4}{*}{\raggedright\arraybackslash UFLP} & 110 & 3381.7 & 3132.8 & 7907290.8 & 7863352.6\\
\cmidrule{1-7}
 &  & 20 & 7.0 & 2.4 & 37614.8 & 9719.4\\

 &  & 30 & 42.0 & 36.0 & 209125.3 & 169072.9\\

 & \multirow{-3}{*}{\raggedright\arraybackslash KP} & 40 & 854.8 & 751.8 & 3278348.2 & 2963984.3\\
 
 \cmidrule{2-7}

 &  & 42 & 558.9 & 306.4 & 965511.4 & 542167.4\\

\multirow{-5}{*}{\raggedleft\arraybackslash 4} & \multirow{-2}{*}{\raggedright\arraybackslash UFLP} & 56 & 3600.0 & 3513.8 & 3288090.2 & 3324261.9\\
\bottomrule
\end{tabu}
        }
        \caption{Average CPU time expressed in seconds and average number of linear programs solved over 10 instances for each problem class, number of objectives, and number of variables. Two algorithms %configurations 
        are tested here: \texttt{BB} and \texttt{BB-E}. %For CFLP, the first value uses the branching rule \texttt{MOF} whereas the second uses \texttt{PS}. %The fastest %configuration 
        %of the three is in bold.
        }
        \label{tab:OSSenum}
    \end{table}
    
    In general, the enumeration procedure reduces the CPU time, even for larger instances. For CFLP, enumeration appears to be slightly slower (e.g., less than $0.5\%$ for $n = 230$). As expected, the proportion of CPU time spent in updating the upper bound set is greater in \texttt{BB-E}, but fewer linear programs are solved.
    
    % ---------------------------------------
    %   Question 4
    % ---------------------------------------
    
    \subsection{Problem specific variable selection rules:} %Investigating variable selection rules: 
    the case of CFLP
    \label{sec:expeVarSel}
    
    By performing probing, the algorithm is able to reduce the set of branching candidates. This indirectly helps the algorithm to make better branching decisions, as variables that lead to redundant branches are discarded. In this section, we study whether a tailored %having a good 
    variable selection rule could help to further improve the CPU time. For this purpose, we limit the analysis to the CFLP, for which a good branching rule is well known in the single objective case: branch first on the variables that handle the opening of the facilities. Note that this rule is valid for UFLP as well, but the preliminary results showed that it had very little impact for UFLP. Indeed, many integer solutions are already found in the early stages of the algorithm, which implies that it is not necessarily possible to reach them much earlier by branching on the facility variables first. Thus, we omitted the analysis for UFLP.
    We limit ourselves to the best known configuration so far, i.e., \texttt{BB-E}. % FOB-VFD-BBWSN
    No cover cuts are generated. Two configurations are tested here: the regular variable selection rule (\texttt{MOF}, for Most Often Fractional), and the problem specific selection rule (\texttt{PS}, for Problem Specific).
    
    \begin{table}[]
        \centering
        \resizebox{\linewidth}{!}{
        
\begin{tabu} to 1.4\linewidth {>{\centering}X>{\centering}X>{\centering}X>{\centering}X>{\centering}X}
\toprule
\multicolumn{1}{c}{ } & \multicolumn{2}{c}{CPU} & \multicolumn{2}{c}{\# unsolved} \\
\cmidrule(l{3pt}r{3pt}){2-3} \cmidrule(l{3pt}r{3pt}){4-5}
n & \texttt{MOF} & \texttt{PS} & \texttt{MOF} & \texttt{PS}\\
\midrule
65 & 10.0 & 12.3 & 0 & 0\\
230 & 1993.3 & 581.4 & 1 & 0\\
495 & 3600.0 & 3600.0 & 10 & 10\\
\bottomrule
\end{tabu}
        }
        \caption{Comparison of a problem specific branching rule \texttt{PS} and the general purpose rule \texttt{MOF} for CFLP. % are shown in this table. 
        For each instance class given by the number of variables $n$, %it reports
        the average CPU time expressed in seconds (CPU) and the number of unsolved instances (\# unsolved) is given.% for two configurations considered, namely \texttt{MOF} and \texttt{PS}.
        }
        \label{tab:CFLP}
    \end{table}
    
    The results presented in Table~\ref{tab:CFLP} clearly show that the problem specific rule (\texttt{PS}) results in important CPU time reductions for medium-sizes instances.  %systematically outperforms the generic rule (\texttt{MOF}). 
    This suggests %impilies
    that there is considerable %much 
    room for improvement in the design of the variable selection rule and future research should address this aspect in more detail.
    %This aspect of MOBB has not received a lot of attention in the literature to this day, and this is left for future research.

    % ---------------------------------------
    %   Question 5
    % ---------------------------------------
    
    \subsection{Comparison with objective space search algorithms}
    
    In this section, we compare our MOBB framework to a state-of-the-art %an
    objective space search algorithm. For this comparison, we use a C\texttt{++} implementation of the redundancy avoidance method of \cite{Klamroth15}. The implementation was kindly shared with us by \cite{daechert22}. The idea is to decompose the objective space based on the local upper bounds, and to explore each sub-region independently while avoiding redundant regions. In that aspect, the algorithm we use is similar to \cite{Tamby20}, except that the implementation we use does not go as far as theirs in the fine-tuning of CPLEX's parameters. 
    
    The OSS algorithm is compared to the best generic configuration of our MOBB, namely \texttt{BB-E}.
    %The best configuration is used for the MOBB, namely \texttt{BB-E}. 
    No cover cuts are generated, and the generic variable selection rule is used for CFLP (\texttt{MOF}). Consequently, two configurations are tested here: our branch-and-bound framework (\texttt{BB-E}) and the objective space search algorithm (\texttt{OSS}). The results are reported in Table~\ref{tab:OSS}.
    
    \begin{table}[]
        \centering
        \resizebox{\linewidth}{!}{
        
\begin{tabu} to 1.4\linewidth {>{\centering}X>{\centering}X>{\centering}X>{\centering}X>{\centering}X}
\toprule
p & pb & n & \texttt{BB-E (BB-E+CB)} & \texttt{OSS}\\
\midrule
 &  & 65 & 10.0 & \textbf{2.3}\\

 &  & 230 & 1993.3 & \textbf{144.8}\\

 & \multirow{-3}{*}{\raggedright\arraybackslash CFLP} & 495 & 3600.0 & \textbf{731.4}\\
 
 \cmidrule{2-5}

 &  & 40 & 42.5 & \textbf{12.1}\\

 &  & 50 & 108.2 & \textbf{24.8}\\

 &  & 60 & 431.0 & \textbf{73.5}\\

 &  & 70 & 1341.1 & \textbf{191.8}\\

 & \multirow{-5}{*}{\raggedright\arraybackslash KP} & 80 & 2388.9 & \textbf{317.1}\\
 
 \cmidrule{2-5}

 &  & 56 & 134.1 (\textbf{124.8}) & 197.7\\

 &  & 72 & 421.3 (\textbf{414.0}) & 545.0\\

 &  & 90 & 1229.2 (\textbf{1175.3}) & 1342.1\\

\multirow{-12}{*}{\raggedleft\arraybackslash 3} & \multirow{-4}{*}{\raggedright\arraybackslash UFLP} & 110 & 3132.8 (2981.37) & \textbf{2965.0}\\
\cmidrule{1-5}
 &  & 20 & \textbf{2.4} & 8.6\\

 &  & 30 & \textbf{36.0} & 36.1\\

 & \multirow{-3}{*}{\raggedright\arraybackslash KP} & 40 & 751.9 & \textbf{320.3}\\

 \cmidrule{2-5}

 &  & 42 & 306.4 (\textbf{292.8}) & 686.5\\

\multirow{-5}{*}{\raggedleft\arraybackslash 4} & \multirow{-2}{*}{\raggedright\arraybackslash UFLP} & 56 & 3513.8 (\textbf{3414.1}) & 3518.9\\
\bottomrule
\end{tabu}
        }
        \caption{Average CPU time expressed in seconds and number of linear programs solved over 10 instances for each problem class, number of objectives, and number of variables. Two algorithms %configurations 
        are tested here: \texttt{BB-E} and \texttt{OSS}. %The fastest %configuration 
        %of the three is in bold.
        For UFLP, the performance of BB-E using cone objective branching (\texttt{BB-E+CB}) is %also
        reported in brackets, as it was shown to be the best configuration for this problem class (see %in particular in
        Figure~\ref{fig:PerformanceProfileVarfix}).
        }
        \label{tab:OSS}
    \end{table}
    
    First, \texttt{BB-E} seems to be competitive with \texttt{OSS} for UFLP, meaning that the first milestone %step 
    towards an efficient branch-and-bound framework for the multi-objective case has been reached. There is, however, a tendency for larger instances to be solved faster by \texttt{OSS}, e.g., for UFLP, $p = 3$, and $n = 110$. This suggests that there are costs that do not occur for small and medium-sized instances but that become a burden for larger instances. In our opinion, this constitutes a direction for future research. Indeed, now that the branch-and-bound algorithm is competitive on medium-sized instances, the next step is to tackle larger problems. Note that in our experiments, for this specific problem class and problem size, some instances are solved faster using \texttt{OSS} whereas other are solved faster using \texttt{BB-E} (in particular with cone bounding enabled).
    
     Regarding CFLP and KP with $p = 3$, \texttt{OSS} is significantly faster than \texttt{BB-E}. We have seen in Section~\ref{sec:expeVarSel} that major improvements could be obtained for CFLP by working on variable selection rules. % in the future. 
    \cite{Gadegaard2019} showed that cut generation at the root node is very effective for bi-objective CFLP, and this may apply to any number of objectives as well as to other problem classes. These two elements constitute very promising %other research 
    directions for future research. %the future.
    
    Finally, it is interesting to note that \texttt{BB-E} is the most efficient on problems for which the number of non-dominated points per variable ($|\YN|/n$) is the highest. Indeed, over all instances solved by at least one of the configurations, UFLP has $102.84$ non-dominated points per variable, whereas KP and CFLP have $18.74$ and $5.13$, respectively. The results from this section confirm the intuition that OSS algorithms are more efficient on problems with fewer non-dominated points, and benefit a lot from the decades of progress embedded in single-objective solvers, which are particularly competitive on problems with a high number of variables.

\section{Conclusion}
\label{sec:conclu}

%In this paper, we enhanced objective branching for MOBB with three or more objectives, and proposed new and simple node selection rules based on the best-bound idea. Furthermore, we studied the performance of other features such as cut generation, problem specific variable selection rules, and enumeration of solution close to the leaf nodes.

%First, our experiments showed that probing, when coupled with objective branching, lead to significantly smaller search trees, resulting in faster CPU times. Second, we showed that we could overcome the difficult arbitrage between depth-first and breadth-first by proposing a simple rule based on the best-bound idea. Indeed, the new rule resulted in better CPU times than both breadth and depth-first strategies on the set of instances considered. Finally, we showed that... comparison with OSS?

In this paper, we first enhanced objective branching for MOBB with three or more objective functions. The experiments showed that combining probing and objective branching leads to significantly smaller search trees, resulting in lower CPU times.

Then, we proposed node selection rules based on the best-bound idea. Two variants were considered, depending on how the quality of a bound is measured: either based on the minimal value of a weighted-sum scalarization or on the smallest gap between upper and lower bound sets. The experiments showed that the former is the most efficient, and performs better than both the traditional depth-first and breadth-first strategies from the literature. Besides, we have observed that computing gaps during the resolution can be expensive. This opens the discussion on appropriate gap measures for multi-objective optimization, in particular for $p \geq 3$, and how to efficiently compute these gaps.

Moreover, other developments on additional features were explored. We first investigated cut generation %tried to generate cuts 
based on the objective branching constraints, but this resulted in slower performances due to the increased complexity of the lower bound sets. %Instead, 
Generating cuts only in the root node may be more beneficial, as done by \cite{Gadegaard2019} for the bi-objective case and constitutes a promising direction for future research. %This is left for future research. 
%Afterward, we 
We also showed in our %the 
experiments that although variable fixing reduces the set of potential branching candidates when doing decision space branching, there are still possible improvements to be achieved by identifying appropriate variables to branch on. This constitute another promising direction for future research. Then, enumeration techniques were tested and generated a speed-up in most cases.

Finally, our branch-and-bound framework is the first that proves to be competitive against a state-of-the-art objective space search algorithm on UFLP instances with three and four objectives. Future research should address the development of appropriate techniques for more efficiently solving large scale problem instances. %A possible step for future research is to develop techniques for solving large scale problems.

\vspace{+5mm}
\noindent\textbf{{\large Acknowledgment}}%

% Enter the text of acknowledgments here
\noindent This research was funded in whole, or in part, by the Austrian Science Fund (FWF) [P 31366]. For the purpose of open access, the author has applied a CC BY public copyright licence to any Author Accepted Manuscript version arising from this submission.

\bibliographystyle{apalike} % plainnat
\bibliography{0main_arxiv}     % Remember to make a local copy of BibTeX file when paper finished (use JabRef)

\begin{thebibliography}{}

\bibitem[Adelgren and Gupte, 2022]{adelgren2021branch}
Adelgren, N. and Gupte, A. (2022).
\newblock Branch-and-bound for biobjective mixed-integer linear programming.
\newblock {\em INFORMS Journal on Computing}, 34(2):909--933.

\bibitem[An et~al., 2022]{DuleabomCFLP}
An, D., Parragh, S.~N., Sinnl, M., and Tricoire, F. (2022).
\newblock A matheuristic for tri-objective binary integer programming.
\newblock Technical report.

\bibitem[Belotti et~al., 2016]{Belotti_2016}
Belotti, P., Soylu, B., and Wiecek, M. (2016).
\newblock Fathoming rules for biobjective mixed integer linear programs: Review
  and extensions.
\newblock {\em Discrete Optimization}, 22:341--363.

\bibitem[Benson, 1998]{Benson1998}
Benson, H.~P. (1998).
\newblock An outer approximation algorithm for genrating all efficient extreme
  points in the outcome set of a multiple objective linear programming problem.
\newblock {\em Journal of Global Optimization}, 13:1--24.

\bibitem[Boland et~al., 2017]{BolandQuadrantShrinking}
Boland, N., Charkhgard, H., and Savelsbergh, M. (2017).
\newblock The quadrant shrinking method: A simple and efficient algorithm for
  solving tri-objective integer programs.
\newblock {\em European Journal of Operational Research}, 260(3):873 -- 885.

\bibitem[Boland and Savelsbergh, 2016]{BolandLShaped}
Boland, N. and Savelsbergh, H. C.~M. (2016).
\newblock The l-shape search method for triobjective integer programming.
\newblock {\em Mathematical Programming Computation}, 8(2):217--251.

\bibitem[D{\"a}chert et~al., 2021]{daechert22}
D{\"a}chert, K., Fleuren, T., and Klamroth, K. (2021).
\newblock A simple, efficient and versatile objective space algorithm for
  multiobjective integer programming.
\newblock working paper.

\bibitem[Ehrgott, 2005]{Ehrgott05}
Ehrgott, M. (2005).
\newblock {\em Multicriteria Optimization}.
\newblock Springer {B}erlin, Heidelberg, 2nd edition.

\bibitem[Ehrgott and Gandibleux, 2007]{Ehrgott07}
Ehrgott, M. and Gandibleux, X. (2007).
\newblock Bound sets for biobjective combinatorial optimization problems.
\newblock {\em Computers \& Operations Research}, 34(9):2674--2694.

\bibitem[Forget et~al., 2020a]{Forget21OB}
Forget, N., Gadegaard, S., Klamroth, K., Nielsen, L., and Przybylski, A.
  (2020a).
\newblock Branch-and-bound and objective branching with three objectives.
\newblock \url{http://www.optimization-online.org/DB_HTML/2020/12/8158.html}.
\newblock {P}reprint.

\bibitem[Forget et~al., 2022]{Forget2022}
Forget, N., Gadegaard, S.~L., and Nielsen, L.~R. (2022).
\newblock Warm-starting lower bound set computations for branch-and-bound
  algorithms for multi objective integer linear programs.
\newblock {\em European Journal of Operational Research}, 302(3):909--924.

\bibitem[Forget et~al., 2020b]{Forget20a}
Forget, N., Nielsen, L., and Gadegaard, S. (2020b).
\newblock Computational results (all instances).
\newblock Technical report, Aarhus University.
\newblock Results for all the instances at the repository MOrepo-Forget20.

\bibitem[Gadegaard et~al., 2019]{Gadegaard2019}
Gadegaard, S., Nielsen, L., and Ehrgott, M. (2019).
\newblock Bi-objective branch-and-cut algorithms based on lp relaxation and
  bound sets.
\newblock {\em INFORMS Journal on Computing}, 31(4):790--804.

\bibitem[Gu et~al., 1998]{gu1998lifted}
Gu, Z., Nemhauser, G.~L., and Savelsbergh, M.~W. (1998).
\newblock Lifted cover inequalities for 0-1 integer programs: Computation.
\newblock {\em INFORMS Journal on Computing}, 10(4):427--437.

\bibitem[Hamel et~al., 2013]{Hamel2013}
Hamel, A.~H., Löhne, A., and Rudloff, B. (2013).
\newblock Benson type algorithms for linear vector optimization and
  applications.
\newblock {\em Journal of Global Optimization}, 59(4):811--836.

\bibitem[Kirlik, 2014]{Moolibrary-Kirlik14}
Kirlik, G. (2014).
\newblock Test instances for multiobjective discrete optimization problems.

\bibitem[Kirlik and Sayın, 2014]{KirlikSayinTwoStageScalarazation}
Kirlik, G. and Sayın, S. (2014).
\newblock A new algorithm for generating all nondominated solutions of
  multiobjective discrete optimization problems.
\newblock {\em European Journal of Operational Research}, 232(3):479 -- 488.

\bibitem[Kiziltan and Yucao\u{g}lu, 1983]{Kiziltan1983}
Kiziltan, G. and Yucao\u{g}lu, E. (1983).
\newblock An algorithm for multiobjective zero-one linear programming.
\newblock {\em Management Science}, 29(12):1444--1453.

\bibitem[Klamroth et~al., 2015]{Klamroth15}
Klamroth, K., Lacour, R., and Vanderpooten, D. (2015).
\newblock On the representation of the search region in multi-objective
  optimization.
\newblock {\em European Journal of Operational Research}, 245:767--778.

\bibitem[Klein and Hannan, 1982]{KleinHannan1982}
Klein, D. and Hannan, E. (1982).
\newblock An algorithm for the multiple objective integer linear programming
  problem.
\newblock {\em European Journal of Operational Research}, 9(4):378 -- 385.

\bibitem[Letchford and Souli, 2019]{Letchford_2019}
Letchford, A.~N. and Souli, G. (2019).
\newblock On lifted cover inequalities: A new lifting procedure with unusual
  properties.
\newblock {\em Operations Research Letters}, 47(2):83--87.

\bibitem[Linderoth and Savelsbergh, 1999]{linderoth1999computational}
Linderoth, J.~T. and Savelsbergh, M.~W. (1999).
\newblock A computational study of search strategies for mixed integer
  programming.
\newblock {\em INFORMS Journal on Computing}, 11(2):173--187.

\bibitem[L{\"o}hne and Wei{\ss}ing, 2020]{BensolveA}
L{\"o}hne, A. and Wei{\ss}ing, B. (2020).
\newblock Bensolve - vlp solver, version 2.1.x.
\newblock \url{http://www.bensolve.org}.

\bibitem[Mavrotas and Diakoulaki, 1998]{mavrotas1998branch}
Mavrotas, G. and Diakoulaki, D. (1998).
\newblock A branch and bound algorithm for mixed zero-one multiple objective
  linear programming.
\newblock {\em European Journal of Operational Research}, 107(3):530--541.

\bibitem[Mavrotas and Diakoulaki, 2005]{mavrotas2005branch}
Mavrotas, G. and Diakoulaki, D. (2005).
\newblock Multi-criteria branch and bound: A vector maximization algorithm for
  mixed 0-1 multiple objective linear programming.
\newblock {\em Applied {M}athematics and {C}omputation}, 171(1):53--71.

\bibitem[Ozlen et~al., 2014]{OzlenRecursion}
Ozlen, M., Burton, B., and MacRae, C. (2014).
\newblock Multi-objective integer programming: An improved recursive algorithm.
\newblock {\em Journal of Optimization Theory and Applications},
  160(2):470--482.

\bibitem[Parragh and Tricoire, 2019]{parragh2019branch}
Parragh, S. and Tricoire, F. (2019).
\newblock Branch-and-bound for bi-objective integer programming.
\newblock {\em INFORMS Journal on Computing}, 31(4):805--822.

\bibitem[Ramos et~al., 1998]{Ramos1998}
Ramos, R.~M., Alonso, S., Sicilia, J., and Gonz{\'a}lez, C. (1998).
\newblock The problem of the optimal biobjective spanning tree.
\newblock {\em European Journal of Operational Research}, 111(3):617 -- 628.

\bibitem[Santis et~al., 2020]{DeSantis2020}
Santis, M.~D., Eichfelder, G., Niebling, J., and Rocktäschel, S. (2020).
\newblock Solving multiobjective mixed integer convex optimization problems.
\newblock {\em SIAM Journal on Optimization}, 30(4):3122--3145.

\bibitem[Savelsbergh, 1994]{savelsbergh1994preprocessing}
Savelsbergh, M.~W. (1994).
\newblock Preprocessing and probing techniques for mixed integer programming
  problems.
\newblock {\em ORSA Journal on Computing}, 6(4):445--454.

\bibitem[Sourd and Spanjaard, 2008]{SourdSpanjaard2008}
Sourd, F. and Spanjaard, O. (2008).
\newblock A multiobjective branch-and-bound framework: Application to the
  biobjective spanning tree problem.
\newblock {\em INFORMS Journal on Computing}, 20(3):472--484.

\bibitem[Stidsen and Andersen, 2018]{Stidsen18}
Stidsen, T. and Andersen, K.~A. (2018).
\newblock A hybrid approach for biobjective optimization.
\newblock {\em Discrete Optimization}, 28:89--114.

\bibitem[Stidsen et~al., 2014]{StidsenAndersenDammann2014}
Stidsen, T., Andersen, K.~A., and Dammann, B. (2014).
\newblock A branch and bound algorithm for a class of biobjective mixed integer
  programs.
\newblock {\em Management Science}, 60(4):1009--1032.

\bibitem[Sylva and Crema, 2004]{SylvaCremaDisjunction}
Sylva, J. and Crema, A. (2004).
\newblock A method for finding the set of non-dominated vectors for multiple
  objective integer linear programs.
\newblock {\em European Journal of Operational Research}, 158(1):46 -- 55.

\bibitem[Tamby and Vanderpooten, 2021]{Tamby20}
Tamby, S. and Vanderpooten, D. (2021).
\newblock Enumeration of the nondominated set of multiobjective discrete
  optimization problems.
\newblock {\em INFORMS Journal on Computing}, 33(1):72--85.

\bibitem[Ulungu and Teghem, 1995]{UlunguTeghem1995}
Ulungu, E. and Teghem, J. (1995).
\newblock The two phases method: An efficient procedure to solve bi-objective
  combinatorial optimization problems.
\newblock {\em Foundations of Computing and Decision Sciences}, 20(2):149--165.

\bibitem[Vincent et~al., 2013]{Vincent2013corrections}
Vincent, T., Seipp, F., Ruzika, S., Przybylski, A., and Gandibleux, X. (2013).
\newblock Multiple objective branch and bound for mixed 0-1 linear programming:
  Corrections and improvements for the biobjective case.
\newblock {\em Computers \& Operations Research}, 40(1):498--509.

\bibitem[Vis{\'e}e et~al., 1995]{Visee95}
Vis{\'e}e, M., Teghem, J., Pirlot, M., and Ulungu, E. (1995).
\newblock The two-phases method: An efficient procedure to solve bi-objective
  combinatorial optimization problems.
\newblock {\em Foundations of Computing and Decision Science}, 20:149--165.

\bibitem[Vis{\'e}e et~al., 1998]{ViseeUlunguTeghem1998}
Vis{\'e}e, M., Teghem, J., Pirlot, M., and Ulungu, E.~L. (1998).
\newblock Two-phases method and branch and bound procedures to solve the
  bi--objective knapsack problem.
\newblock {\em Journal of Global Optimization}, 12(2):139--155.

\end{thebibliography}

\appendix
\appendixpage
%\newpage
%\section{Problem class}
\section{Considered benchmark problems}
    \subsection{CFLP}
    
    In this problem, there is a set of $l$ locations where a facility can be opened, and a set of $r$ customers that each have to be assigned to a location. Two decisions have to be made: which locations to open and which customers to assign to which facilities. Both opening a facility and assigning a customer to an open facility induces a cost. Moreover, the demand being handled in a facility $j$ cannot exceed a threshold $t_j$. Finally, the company may chose to ignore some of the customers to reduce their costs.
    
    The first objective is to minimize the cost of assigning customers to facilities. The second objective consists in minimizing the opening cost of the facilities. The third objective aims to maximize the overall demand of the customers that is satisfied.
    
    Let $y_j = 1$ if a facility is opened at location $j$, and $y_j = 0$ otherwise, $\forall j \in \{1,...,l\}$. Furthermore, let $x_{ij} = 1$ if customer $i$ is assigned to location $j$, and $x_{ij} = 0$ otherwise, $\forall i \in \{1,...,r\}, \forall j \in \{1,...,l\}$. Finally, $z_i = 1$ if customer $i$ is served, $0$ otherwise, $\forall i \in \{1,...,r\}$.
    
\begin{align}
	\min        \ &\sum_{i=1}^r\sum_{j=1}^lc_{ij}x_{ij}&&\\
	\min \ & \sum_{j=1}^lf_jy_j && \\
	\max \ & \sum_{i=1}^r d_iz_i && \\
	\mbox{s.t.}\ & \sum_{j=1}^l x_{ij}=z_i&&\forall i\in\{1,\dots,r\}\\
	\ & x_{ij}\leq y_j&&\forall i\in\{1,\dots,r\},\ j\in\{1,\dots,l\}\\
	\ & \sum_{i=1}^r d_ix_{ij} \leq t_jy_j && \forall j\in\{1,\dots,l\}\\
	\ & x_{ij}\in\{0,1\} &&\forall i\in\{1,\dots,r\},\ j\in\{1,\dots,l\}\\
	\ & y_j\in\{0,1\}&&\forall j\in\{1,\dots,l\}\\
	\ & z_i \in \{0,1\}&&\forall i\in\{1,\dots,r\}
\end{align}

    %The demands $d_i$ are generated randomly in the range $[5,10]$.
Instances are taken from \cite{DuleabomCFLP}.

    \subsection{KP}
    
In the Knapsack problem, a subset of items has to be selected from a set of $n$ items. Each item $i$ has a weight $w_i$, and there is a limit $b$ on the total weight of the items being selected. Moreover, each item $i$ has a utility $c_i^k$ in objective $k$, and the goal is to maximize the utility of the subset of items selected over all objective functions.

Let $x_i = 1$ if item $i$ is selected, $0$ otherwise. The multi-objective Knapsack Problem (KP) with $p$ objectives can be formulated as follows:

\begin{align}
	\min        \ &\sum_{i=1}^nc_i^kx_i&&\forall k\in\{1,...,p\}\\
	\mbox{s.t.}\ & \sum_{i=1}^n w_i x_i \leq b&&\\
	\ & x_i\in\{0,1\}&&\forall i\in\{1,\dots,n\}
\end{align}

Instances are taken from \cite{KirlikSayinTwoStageScalarazation}.
%All $c_i^k$ and $w_i$ are generated randomly in the range $[1,1000]$, and $b$ is set to $\frac{1}{2} \sum_{i=1}^n w_i$

    \subsection{UFLP}
    
In this problem, there is a set of $l$ locations where a facility can be opened, and a set of $r$ customers that each have to be assigned to a location. Two decisions have to be made: which locations to open and which customers to assign to which facilities. Both opening a facility and assigning a customer to an open facility induces a cost, and the overall cost has to be minimized. Let $y_j = 1$ if a facility is opened at location $j$, and $y_j = 0$ otherwise, $\forall j \in \{1,...,l\}$. Furthermore, let $x_{ij} = 1$ if customer $i$ is assigned to location $j$, and $x_{ij} = 0$ otherwise, $\forall i \in \{1,...,r\}, \forall j \in \{1,...,l\}$.

The multi-objective Uncapacitated Facility Location Problem (UFLP) with $p$ objectives can be formulated as the following MOCO problem

\begin{align}
	\min        \ &\sum_{i=1}^r\sum_{j=1}^lc_{ij}^kx_{ij}+\sum_{j=1}^lf_j^ky_j&&\forall k\in\{1,...,p\}\\
	\mbox{s.t.}\ & \sum_{j=1}^l x_{ij}=1&&\forall i\in\{1,\dots,r\}\\
	\ & x_{ij}\leq y_j&&\forall i\in\{1,\dots,r\},\ j\in\{1,\dots,l\}\\
	\ & x_{ij}\in\{0,1\} &&\forall i\in\{1,\dots,r\},\ j\in\{1,\dots,l\}\\
	\ & y_j\in\{0,1\}&&\forall j\in\{1,\dots,l\}
\end{align}

%The cost for assigning customer $i$ to facility $j$ in objective $k$ is given by $c_{ij}^k$. The assignment costs belongs to the interval $[1,1000]$ and is generated on the non-dominated part (in minimization) of a hypersphere of dimension $p$ (see \cite{gMOIP} for further details). 
%The cost for opening a facility on location $j$ is given by $f_j$. The fixed opening costs are generated from the interval $[1,100]$ and, like the assignment costs, these coefficients are generated on the non-dominated part (in minimization) of a hypersphere of dimension $p$. The number of variables in this problem is $n = l(r+1)$.
Instances are taken from \cite{Forget2022}.
    
\newpage
%\section{Table \texttt{BBGAP}}
\section{Additional computational results}

\begin{table}[H]
        \centering
        %\resizebox{!}{0.25\textwidth}{\input{tablesExpe/tabVarFixNodes}}
        %\begin{table}
%\centering
\resizebox{0.95\linewidth}{!}{
\begin{tabular}{lrrrrrrrr}
\toprule
\multicolumn{3}{c}{ } & \multicolumn{2}{c}{\texttt{CB}} & \multicolumn{2}{c}{	\texttt{FOB}} & \multicolumn{2}{c}{\texttt{NOB}} \\
\cmidrule(l{3pt}r{3pt}){4-5} \cmidrule(l{3pt}r{3pt}){6-7} \cmidrule(l{3pt}r{3pt}){8-9}
p & pb & n & \texttt{NVF} & \texttt{VF} & \texttt{NVF} & \texttt{VF} & \texttt{NVF} & \texttt{VF}\\
\midrule
 &  & 65 & 7906.6  ( 0 ) & 2382.6  ( 0 ) & 11786.2  ( 0 ) & 1389.4  ( 0 ) & 15090.4  ( 0 ) & 15072.4  ( 0 )\\

 &  & 230 & 198405.7  ( 6 ) & 65681.8  ( 4 ) & 405597.4  ( 7 ) & 49365.9  ( 1 ) & 78613.1  ( 10 ) & 32593.7  ( 10 )\\

 & \multirow{-3}{*}{\raggedright\arraybackslash CFLP} & 495 & 6380.6  ( 10 ) & 4357.9  ( 10 ) & 33545.9  ( 10 ) & 4497.9  ( 10 ) & 4245.2  ( 10 ) & 2934.4  ( 10 )\\
 
 \cmidrule{3-9}

 &  & 40 & 104268.4  ( 0 ) & 15918.4  ( 0 ) & 136980.2  ( 0 ) & 13023.6  ( 0 ) & 137544.2  ( 0 ) & 71502.4  ( 0 )\\

 &  & 50 & 283955  ( 0 ) & 34498.6  ( 0 ) & 385177  ( 0 ) & 27261.8  ( 0 ) & 385625.2  ( 0 ) & 174757.2  ( 0 )\\

 &  & 60 & 804121.5  ( 1 ) & 121213.2  ( 0 ) & 1128878.7  ( 2 ) & 95023.4  ( 0 ) & 623624.4  ( 9 ) & 428664.6  ( 7 )\\

 &  & 70 & 1310720.2  ( 7 ) & 289369.3  ( 3 ) & 1839920.2  ( 7 ) & 226918  ( 2 ) & 617265.9  ( 10 ) & 456791.8  ( 10 )\\

 & \multirow{-5}{*}{\raggedright\arraybackslash KP} & 80 & 1368612.3  ( 10 ) & 377936.3  ( 5 ) & 1812818.1  ( 10 ) & 303356  ( 5 ) & 499527.9  ( 10 ) & 367138.9  ( 10 )\\
 
 \cmidrule{3-9}

 &  & 56 & 49212.2  ( 0 ) & 20885.2  ( 0 ) & 216010.6  ( 0 ) & 14512.6  ( 0 ) & 63239  ( 0 ) & 62944.2  ( 0 )\\

 &  & 72 & 111566.6  ( 0 ) & 47018.2  ( 0 ) & 531514  ( 0 ) & 33598.6  ( 0 ) & 151994.2  ( 0 ) & 151525.4  ( 0 )\\

 &  & 90 & 229470.8  ( 0 ) & 97638.6  ( 0 ) & 1103596  ( 5 ) & 72544.4  ( 0 ) & 275612.6  ( 6 ) & 235776.7  ( 9 )\\

\multirow{-12}{*}{\raggedleft\arraybackslash 3} & \multirow{-4}{*}{\raggedright\arraybackslash UFLP} & 110 & 195819.1  ( 10 ) & 157547.9  ( 5 ) & 681660.1  ( 10 ) & 73505.6  ( 10 ) & 71013.9  ( 10 ) & 72963.3  ( 10 )\\
\cmidrule{1-9}
 &  & 20 & 7651.6  ( 0 ) & 2103.6  ( 0 ) & 8442.4  ( 0 ) & 1899.4  ( 0 ) & 8998  ( 0 ) & 5043.2  ( 0 )\\

 &  & 30 & 43529.8  ( 0 ) & 8469.2  ( 0 ) & 49641  ( 0 ) & 7589.4  ( 0 ) & 53294.2  ( 0 ) & 26693.4  ( 0 )\\

 & \multirow{-3}{*}{\raggedright\arraybackslash KP} & 40 & 330758.7  ( 2 ) & 82242.8  ( 2 ) & 430534.1  ( 2 ) & 72322.6  ( 1 ) & 340657.3  ( 7 ) & 221146.5  ( 3 )\\
 
 \cmidrule{3-9}

 &  & 42 & 62617.8  ( 0 ) & 29925.4  ( 0 ) & 199163  ( 0 ) & 22728.8  ( 0 ) & 68111.2  ( 0 ) & 68070.2  ( 0 )\\

\multirow{-5}{*}{\raggedleft\arraybackslash 4} & \multirow{-2}{*}{\raggedright\arraybackslash UFLP} & 56 & 71905  ( 10 ) & 49605.7  ( 10 ) & 172122.8  ( 10 ) & 32013.5  ( 10 ) & 183316.4  ( 10 ) & 100648.1  ( 10 )\\
\bottomrule
\end{tabular}}
%\end{table}
        \caption{The average number of nodes explored over 10 instances for each problem class, number of objectives, number of variables, and configuration. The number in brackets is the number of instances unsolved. Note that when the number of unsolved instances is high, the number of nodes explored may be low due to the fact that the algorithm could not explore a large number of nodes within the time limit of one hour.}
        \label{tab:varFixNodes}
    \end{table}

\begin{table}
    \centering
    \resizebox{\linewidth}{!}{
    
\begin{tabu} to 1.4\linewidth {>{\raggedleft}X>{\raggedright}X>{\raggedleft}X>{\raggedleft}X>{\raggedleft}X>{\raggedleft}X>{\raggedleft}X>{\raggedleft}X>{\raggedleft}X>{\raggedleft}X>{\raggedleft}X>{\raggedleft}X>{\raggedleft}X>{\raggedleft}X>{\raggedleft}X>{\raggedleft}X>{\raggedleft}X>{\raggedleft}X>{\raggedleft}X>{\raggedleft}X>{\raggedleft}X}
\toprule
\multicolumn{3}{c}{ } & \multicolumn{6}{c}{\% LB set} & \multicolumn{6}{c}{\% Probing} & \multicolumn{6}{c}{\% Other} \\
\cmidrule(l{3pt}r{3pt}){4-9} \cmidrule(l{3pt}r{3pt}){10-15} \cmidrule(l{3pt}r{3pt}){16-21}
\multicolumn{3}{c}{ } & \multicolumn{2}{c}{\texttt{CB}} & \multicolumn{2}{c}{\texttt{FOB}} & \multicolumn{2}{c}{\texttt{NOB}} & \multicolumn{2}{c}{\texttt{CB}} & \multicolumn{2}{c}{\texttt{FOB}} & \multicolumn{2}{c}{\texttt{NOB}} & \multicolumn{2}{c}{\texttt{CB}} & \multicolumn{2}{c}{\texttt{FOB}} & \multicolumn{2}{c}{\texttt{NOB}} \\
\cmidrule(l{3pt}r{3pt}){4-5} \cmidrule(l{3pt}r{3pt}){6-7} \cmidrule(l{3pt}r{3pt}){8-9} \cmidrule(l{3pt}r{3pt}){10-11} \cmidrule(l{3pt}r{3pt}){12-13} \cmidrule(l{3pt}r{3pt}){14-15} \cmidrule(l{3pt}r{3pt}){16-17} \cmidrule(l{3pt}r{3pt}){18-19} \cmidrule(l{3pt}r{3pt}){20-21}
p & pb & n & \texttt{NVF} & \texttt{VF} & \texttt{NVF} & \texttt{VF} & \texttt{NVF} & \texttt{VF} & \texttt{NVF} & \texttt{VF} & \texttt{NVF} & \texttt{VF} & \texttt{NVF} & \texttt{VF} & \texttt{NVF} & \texttt{VF} & \texttt{NVF} & \texttt{VF} & \texttt{NVF} & \texttt{VF}\\
\midrule
 &  & 65 & 90.0 & 55.1 & 86.8 & 43.2 & 89.3 & 43.6 & 0.0 & 41.0 & 0.0 & 53.4 & 0.0 & 51.4 & 10.0 & 3.9 & 13.2 & 3.4 & 10.7 & 5.0\\

 &  & 230 & 88.5 & 48.4 & 80.8 & 32.4 & 95.7 & 47.1 & 0.0 & 48.3 & 0.0 & 64.8 & 0.0 & 51.1 & 11.5 & 3.3 & 19.2 & 2.8 & 4.3 & 1.8\\

 & \multirow{-3}{*}{\raggedright\arraybackslash CFLP} & 495 & 98.3 & 73.3 & 93.3 & 53.2 & 98.7 & 70.0 & 0.0 & 25.5 & 0.0 & 45.8 & 0.0 & 29.0 & 1.7 & 1.2 & 6.7 & 1.1 & 1.3 & 0.9\\

 &  & 40 & 90.6 & 69.9 & 88.3 & 62.5 & 95.0 & 68.5 & 0.0 & 25.6 & 0.0 & 32.8 & 0.0 & 28.0 & 9.4 & 4.5 & 11.7 & 4.7 & 5.0 & 3.5\\

 &  & 50 & 89.1 & 68.3 & 86.5 & 59.8 & 95.0 & 68.5 & 0.0 & 26.8 & 0.0 & 35.2 & 0.0 & 28.0 & 10.9 & 4.9 & 13.5 & 5.1 & 5.0 & 3.5\\

 &  & 60 & 87.5 & 66.3 & 84.0 & 56.5 & 94.8 & 66.2 & 0.0 & 27.8 & 0.0 & 37.7 & 0.0 & 29.8 & 12.5 & 5.9 & 16.0 & 5.9 & 5.2 & 3.9\\

 &  & 70 & 85.8 & 63.2 & 81.4 & 50.6 & 94.3 & 59.2 & 0.0 & 29.8 & 0.0 & 42.6 & 0.0 & 36.4 & 14.2 & 7.0 & 18.6 & 6.8 & 5.7 & 4.4\\

 & \multirow{-5}{*}{\raggedright\arraybackslash KP} & 80 & 83.7 & 62.0 & 79.5 & 49.2 & 94.3 & 62.0 & 0.0 & 30.4 & 0.0 & 42.6 & 0.0 & 33.7 & 16.3 & 7.6 & 20.5 & 8.2 & 5.7 & 4.2\\

 &  & 56 & 78.1 & 62.4 & 69.7 & 35.0 & 83.5 & 59.8 & 0.0 & 21.6 & 0.0 & 54.1 & 0.0 & 26.6 & 21.9 & 16.1 & 30.3 & 11.0 & 16.5 & 13.6\\

 &  & 72 & 72.7 & 57.4 & 67.2 & 29.1 & 87.2 & 61.7 & 0.0 & 22.6 & 0.0 & 57.9 & 0.0 & 24.2 & 27.3 & 19.9 & 32.8 & 13.1 & 12.8 & 14.1\\

 &  & 90 & 75.6 & 57.7 & 58.2 & 24.9 & 84.7 & 60.8 & 0.0 & 21.8 & 0.0 & 61.1 & 0.0 & 23.1 & 24.4 & 20.5 & 41.8 & 14.0 & 15.3 & 16.1\\

\multirow{-12}{*}{\raggedleft\arraybackslash 3} & \multirow{-4}{*}{\raggedright\arraybackslash UFLP} & 110 & 63.8 & 51.1 & 47.1 & 18.5 & 84.7 & 66.9 & 0.0 & 21.9 & 0.0 & 66.6 & 0.0 & 17.9 & 36.2 & 27.0 & 52.9 & 14.9 & 15.3 & 15.2\\
\cmidrule{1-21}
 &  & 20 & 93.8 & 85.9 & 93.0 & 84.3 & 95.8 & 82.1 & 0.0 & 9.2 & 0.0 & 10.8 & 0.0 & 14.7 & 6.2 & 4.9 & 7.0 & 4.9 & 4.2 & 3.2\\

 &  & 30 & 91.6 & 82.3 & 90.3 & 79.8 & 95.7 & 81.8 & 0.0 & 10.8 & 0.0 & 13.0 & 0.0 & 14.7 & 8.4 & 6.9 & 9.7 & 7.2 & 4.3 & 3.5\\

 & \multirow{-3}{*}{\raggedright\arraybackslash KP} & 40 & 85.7 & 78.2 & 83.3 & 73.7 & 94.5 & 82.8 & 0.0 & 8.9 & 0.0 & 11.7 & 0.0 & 11.9 & 14.3 & 13.0 & 16.7 & 14.6 & 5.5 & 5.3\\

 &  & 42 & 53.6 & 41.1 & 51.5 & 35.4 & 70.5 & 43.0 & 0.0 & 15.0 & 0.0 & 26.0 & 0.0 & 32.4 & 46.4 & 43.9 & 48.5 & 38.6 & 29.5 & 24.6\\

\multirow{-5}{*}{\raggedleft\arraybackslash 4} & \multirow{-2}{*}{\raggedright\arraybackslash UFLP} & 56 & 36.2 & 29.3 & 30.1 & 20.1 & 50.6 & 33.2 & 0.0 & 12.6 & 0.0 & 30.2 & 0.0 & 35.2 & 63.8 & 58.1 & 69.9 & 49.7 & 49.4 & 31.5\\
\bottomrule
\end{tabu}
    }
    \caption{Comparison of different objective branching settings (cone branching (\texttt{CB}), full objective branching (\texttt{FOB}), no objective branching (\texttt{NOB})) in combination with (\texttt{VF}) and without (\texttt{NVF}) probing concerning the average percentage of CPU time %average proportion 
    spent in different parts of the algorithm over 10 instances for each problem class, number of objectives, and number of variables. %, expressed in percentages of total CPU time. 
    {\it \% LB set} represents the share of CPU time spent in the computation of lower bound sets. {\it \% Probing} represents the proportion of CPU time dedicated to probing. {\it \% Other} is the percentage of CPU time spent in other parts of the algorithms such as dominance test, creation of sub-problems, node selection, etc.}
    \label{tab:apdxProp}
\end{table}

\begin{table}[h]
    \centering
    \resizebox{\linewidth}{!}{
    
\begin{tabu} to 1.4\linewidth {>{\centering}X>{\centering}X>{\centering}X>{\centering}X>{\centering}X>{\centering}X>{\centering}X>{\centering}X}
\toprule
\multicolumn{3}{c}{ } & \multicolumn{2}{c}{CPU} & \multicolumn{2}{c}{\# Nodes} & \multicolumn{1}{c}{ } \\
\cmidrule(l{3pt}r{3pt}){4-5} \cmidrule(l{3pt}r{3pt}){6-7}
p & pb & n & \texttt{BBGAP} & \texttt{BBWSN} & \texttt{BBGAP} & \texttt{BBWSN} & \% CPU \texttt{BBGAP}\\
\midrule
 &  & 65 & 19.6 & 10.0 & 2064.6  ( 0 ) & 1278.8  ( 0 ) & 1.8\\

 &  & 230 & 3718.7 & 1988.0 & 34494.4  ( 10 ) & 40765.4  ( 1 ) & 3.9\\

 & \multirow{-3}{*}{\raggedright\arraybackslash CFLP} & 495 & 3602.3 & 3601.7 & 1918.1  ( 10 ) & 4496.5  ( 10 ) & 5.1\\
 
 \cmidrule{2-8}

 &  & 40 & 124.8 & 46.7 & 5826.8  ( 0 ) & 7405.8  ( 0 ) & 9.8\\

 &  & 50 & 286.5 & 116.5 & 12459.3  ( 0 ) & 14972  ( 0 ) & 12.9\\

 &  & 60 & 1034.9 & 457.8 & 33760  ( 0 ) & 43623  ( 0 ) & 22.4\\

 &  & 70 & 2348.9 & 1414.1 & 57832.4  ( 4 ) & 105508.2  ( 0 ) & 22.2\\

 & \multirow{-5}{*}{\raggedright\arraybackslash KP} & 80 & 3260.9 & 2491.5 & 66489.8  ( 5 ) & 154884.5  ( 3 ) & 21.1\\
 
 \cmidrule{2-8}

 &  & 56 & 235.5 & 177.3 & 7524.1  ( 0 ) & 12282.8  ( 0 ) & 22.6\\

 &  & 72 & 791.1 & 564.4 & 15755.7  ( 0 ) & 26574.4  ( 0 ) & 25.8\\

 &  & 90 & 2539.7 & 1590.5 & 30022.4  ( 0 ) & 52141.6  ( 0 ) & 26.8\\

\multirow{-12}{*}{\raggedleft\arraybackslash 3} & \multirow{-4}{*}{\raggedright\arraybackslash UFLP} & 110 & 3601.1 & 3381.7 & 17823.2  ( 10 ) & 80985.7  ( 5 ) & 34.2\\
\cmidrule{1-8}
 &  & 20 & 35.9 & 7.0 & 923.5  ( 0 ) & 1325.6  ( 0 ) & 8.8\\

 &  & 30 & 150.5 & 42.0 & 3640.7  ( 0 ) & 5105.6  ( 0 ) & 25.8\\

 & \multirow{-3}{*}{\raggedright\arraybackslash KP} & 40 & 2176.1 & 854.8 & 14339.3  ( 4 ) & 41642  ( 0 ) & 51.0\\
 
 \cmidrule{2-8}

 &  & 42 & 3530.1 & 558.9 & 7984.4  ( 7 ) & 22393.8  ( 0 ) & 48.3\\

\multirow{-5}{*}{\raggedleft\arraybackslash 4} & \multirow{-2}{*}{\raggedright\arraybackslash UFLP} & 56 & 3600.0 & 3600.0 & 350.25  ( 10 ) & 39079.1  ( 10 ) & 50.8\\
\bottomrule
\end{tabu}
    }
    \caption{Comparison of node selection rules \texttt{BBWSN} and \texttt{BBGAP}. Columns {\it CPU} give the average CPU time expressed in seconds over 10 instances for each problem class, number of objectives, and number of variables. Columns {\it \# Nodes} provide the average number of nodes explored as well as the number of unsolved instances (indicated in brackets). % are reported for configurations \texttt{BBGAP} and \texttt{BBWSN} respectively. 
    Finally, Column {\it \% CPU} represents the percentage of the total CPU time spent in updating gaps in configuration \texttt{BBGAP}
    }
    \label{tab:apdxGap}
\end{table}

\textcolor{white}{hh}

%\section{CPU times variable fixing}

%    \begin{table}[]
%        \centering
%        \resizebox{!}{0.3\textwidth}{\input{tablesExpe/tabVarFixCPU}}
%        \caption{Average CPU time over 10 instances for each problem class, number of objective, number of variables, and configuration. The time limit is set to one hour.}
%        \label{tab:varFixCPU}
%    \end{table}
%\input{pbExpe}

\end{document}